\PassOptionsToPackage{table}{xcolor}  
\documentclass[letterpaper]{article} 
\usepackage{aaai2027}
\usepackage{times}  
\usepackage{helvet}  
\usepackage{courier}  
\usepackage[hyphens]{url}  
\usepackage{graphicx} 
\usepackage{natbib}  
\usepackage{caption} 
\usepackage{threeparttable}  
\usepackage{tabularx}        

\usepackage{booktabs}        
\usepackage{amssymb}
\usepackage{amsmath}
\usepackage{algorithm}
\usepackage{algpseudocode}

\usepackage{xspace}
\usepackage{amsmath}
\usepackage{amssymb}
\usepackage{multirow} 

\usepackage{subfiles}
\usepackage{booktabs}
\usepackage{multirow}
\usepackage{subcaption}
\usepackage{newfloat}
\usepackage{listings}
\DeclareCaptionStyle{ruled}{labelfont=normalfont,labelsep=colon,strut=off} 
\floatstyle{ruled}
\newfloat{listing}{tb}{lst}{}
\floatname{listing}{Listing}
\newcommand{\method}{EchoEdit\xspace}
\newcommand{\methodplus}{EchoEdit+\xspace}
\newcommand{\nmax}{n_{\max}}
\newcommand{\nmin}{n_{\min}}
\newcommand{\navg}{n_{\mathrm{avg}}}
\newcommand{\vect}[1]{\boldsymbol{#1}}
\newcommand{\src}{\mathrm{src}}
\newcommand{\tar}{\mathrm{tar}}

\nocopyright 

\title{\method: Stabilizing Inversion-Free Audio Editing via Optimal Transport Geometry}
\author{
    Zhongyuan Fu\textsuperscript{\rm 1},
    Yuhang Jia\textsuperscript{\rm 1},
    Hui Wang\textsuperscript{\rm 1},
    Pengjun Chen\textsuperscript{\rm 1},
    Jian Gao\textsuperscript{\rm 1},
    Cun Liu\textsuperscript{\rm 1}, \\
    Wenjia Zeng\textsuperscript{\rm 2},
    Yong Chen\textsuperscript{\rm 2},
    Yong Qin\textsuperscript{\rm 1}\thanks{Corresponding author.},
}
\affiliations{
    \textsuperscript{\rm 1}College of Computer Science, Nankai University, Tianjin, China\\
   \textsuperscript{\rm 2}Lingxi (Beijing) Technology Co., Ltd., Beijing, China \\
    qinyong@nankai.edu.cn
}

\usepackage{bibentry}

\begin{document}

\maketitle

\begin{abstract}
Text-guided audio editing with pretrained generative models is commonly implemented through inversion or noising.
This topology induces a structural trade-off, as stronger edits require deeper corruption of the very rhythm, transients, timbre, and long-range form that should remain unchanged.
Here, we introduce \method, a training-free and inversion-free framework for real-audio editing that directly constructs an editing field by differencing the drifts conditioned on the source and target prompts.
This construction avoids explicit source inversion, paired edit data, and test-time optimization, but its stochastic source marginals introduce uncertainty drift, where small random deviations accumulate along the editing trajectory and can move the edited latent away from the audio data manifold.
To address this limitation, we further propose \methodplus, an optimal-transport-regularized extension that stabilizes the direct editing path by minimizing the transportation cost between edited variables and the source-conditioned audio manifold.
The resulting OT coupling contracts the stochastic displacement at noisy states, keeps model queries closer to the training distribution, and preserves structural information while allowing semantic change.
Experiments on sound-effect and music editing demonstrate that \methodplus improves target-prompt alignment and source preservation over inversion-based baselines and the unregularized direct editor. Code and dataset will be released.
\end{abstract}



\section{Introduction}

Text-conditioned audio generation has recently achieved remarkable progress, enabling the synthesis of diverse audio content, including music, sound effects, and environmental recordings, from natural language descriptions.
Recent systems such as AudioLDM, AudioLDM 2, Tango, Make-An-Audio, AudioGen, MusicGen, and Stable Audio have demonstrated that large-scale pretrained generative models can effectively capture semantic relationships between language and complex audio structures~\cite{liu2023audioldmtexttoaudiogenerationlatent,liu2023audioldm2,ghosal2023texttoaudiogenerationusinginstructiontuned,huang2023makeanaudio,kreuk2023audiogentextuallyguidedaudio,copet2024simplecontrollablemusicgeneration,evans2026stableaudio3}.
For creative workflows, however, transforming existing audio is often more valuable than generating entirely new content, motivating the need for more stable audio editing methods.

Most existing text-guided audio editors inherit an inversion-based topology.
They first move the source audio toward a noisy latent state, either by explicit ODE/DDPM inversion or by direct stochastic noising, and then regenerate the sample under a target prompt~\cite{meng2022sdeditguidedimagesynthesis,manor2024zeroshotunsupervisedtextbasedaudio,zhang2024musicmaguszeroshottexttomusicediting,evans2026stableaudio3}.
This source-to-noise-to-target route creates a fundamental edit-strength trade-off.
\begin{figure}[t]
    \centering
    \includegraphics[width=\columnwidth]{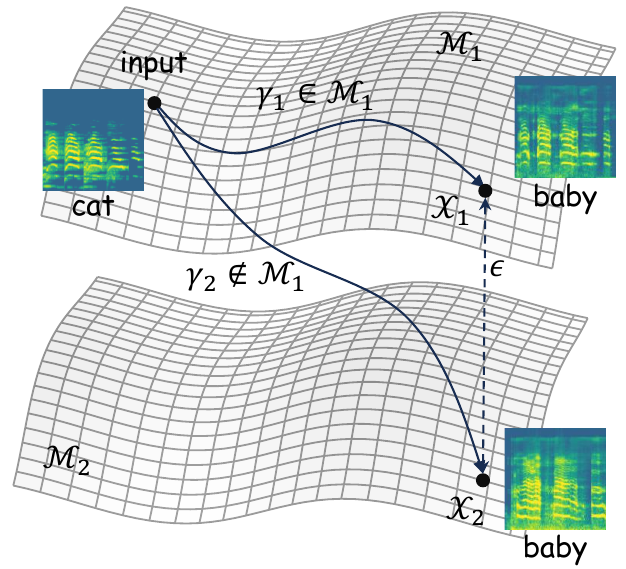}
    \caption{\textbf{The Source-Conditioned Manifold Assumption for Audio Editing.}
    We regard source-preserving editing as transport constrained by a source-conditioned audio manifold $\mathcal{M}_1$, which encodes timing, transient placement, timbre, and long-range structure of the input.
    Although $x_1$ and $x_2$ may both be compatible with the target semantics, they are not equivalent: a trajectory $\gamma_1\in\mathcal{M}_1$ yields a target edit that remains structurally tied to the source, whereas an off-manifold trajectory $\gamma_2\notin\mathcal{M}_1$ may reach a plausible but inconsistent target, reflected by the deviation $\epsilon$.}
    \label{fig:Method}
\end{figure}
Shallow inversion keeps source structure but leaves little room for semantic change, whereas deep inversion allows stronger prompt adherence but discards the fine temporal and spectral cues needed for faithful audio preservation.
Improving the inversion solver can reduce numerical error, but it cannot address the fundamental limitation of high-noise editing because the generation process still starts from a representation where source information has already been degraded.

We therefore ask whether audio editing can be formulated without source inversion.
Our first contribution, \method, answers this question by constructing a direct source-to-target transport for pretrained rectified-flow audio models.
Instead of inverting a source recording into noise, \method queries the source- and target-conditioned velocity fields at matched stochastic audio marginals and integrates their difference as an editing field.
This yields a training-free, paired-data-free, and architecture-agnostic editor that operates directly in the latent space of a pretrained audio flow model.
Conceptually, \method replaces the two-stage inversion path with a differential transport path, where the shared generative dynamics between the two prompts are canceled and the residual field captures the desired semantic edit.

The direct construction, however, reveals a second and more subtle difficulty.
A source recording defines not only an acoustic class but also a local manifold of timing, transient structure, texture, and long-range organization.
For the same target prompt, many target-compatible latents may exist; faithful editing should move from the source to a target-compatible point that remains on the source-conditioned manifold, rather than to a semantically plausible but structurally unrelated point on another manifold (see Figure~\ref{fig:Method}).

Because \method estimates its edit direction from stochastic source marginals, the edited trajectory is not deterministic even when the source and target prompts are fixed.
The randomness is useful for probing the conditional flow, but its accumulated effect can create uncertainty drift: the edited state gradually departs from the source-conditioned audio manifold and the model is queried at latents that are no longer consistent with the rectified-flow training distribution.
For audio, this drift is especially harmful.
An off-manifold deviation may appear as a small latent perturbation, yet it can disrupt transient alignment, phrase-level timing, harmonic continuity, or background ambience across the edited clip.

To close this gap, we introduce \methodplus, an optimal-transport-stabilized extension of \method.
The key idea is to regularize the direct editing path so that the edited variables remain as close as possible to the original audio manifold while still moving toward the target prompt.
We formulate this requirement as an OT problem: among admissible couplings between source-conditioned and edited variables, choose the coupling that minimizes transportation cost along the rectified-flow path.
This coupling contracts the stochastic displacement at noisy states and makes the edited trajectory converge back toward the data manifold rather than drifting away from it.
The resulting method preserves the structural information of the original audio while retaining the inversion-free semantic control of \method.

Our contributions are summarized as follows:
\begin{itemize}
    \item We introduce \method, the first training-free and inversion-free framework for text-guided audio editing with pretrained rectified-flow audio models.
    \item We identify uncertainty drift as a structural limitation of stochastic velocity-difference editing, showing why randomness can compromise consistency in audio even when the expected edit direction is meaningful.
    \item We propose \methodplus, an OT-enhanced editor that minimizes transportation cost between edited variables and the source-conditioned audio manifold, yielding a stable direct editing trajectory.
\end{itemize}

\section{Related Work}
\label{sec:relatedwork}

We provide a full discussion of related work in the supplementary material.
Text-guided audio editing builds on large-scale prompt-conditioned audio generators, including autoregressive models, latent diffusion systems, and rectified-flow backbones~\cite{kreuk2023audiogentextuallyguidedaudio,copet2024simplecontrollablemusicgeneration,liu2023audioldmtexttoaudiogenerationlatent,liu2023audioldm2,ghosal2023texttoaudiogenerationusinginstructiontuned,evans2026stableaudio3,hung2024tangoflux,fei2024fluxmusic}. 
While these models provide strong generative priors, they cannot directly modify existing recordings while preserving temporal structure, timbre, rhythm, and musical form. 
Instruction-tuned editors such as AUDIT, InstructME, EditGen, and recent audio-language editing systems address this limitation through edit supervision~\cite{wang2023auditaudioeditingfollowing,han2023instructmeinstructionguidedmusic,sioros2025editgenharnessingcrossattentioncontrol,ungersbock2025saoinstructfreeformaudioediting,lan2025guidingaudioeditingaudio,tao2026mmeditunifiedframeworkmultitype}.

Most zero-shot audio editors rely on inversion-based pipelines, where the source is first corrupted and then reconstructed under target conditions~\cite{meng2022sdeditguidedimagesynthesis,manor2024zeroshotunsupervisedtextbasedaudio,zhang2024musicmaguszeroshottexttomusicediting,novack2024dittodiffusioninferencetimetoptimization,xu2024promptguidedpreciseaudioediting,jia2025audioeditor}. 
Although effective, such source-to-noise-to-target schemes can degrade audio-specific structures, including transients, phase coherence, rhythmic details, and long-term continuity. 
Flow-based approaches instead model direct source-to-target transport through conditional velocity fields~\cite{gao2026rfmeditingrectifiedflowmatching,kulikov2025floweditinversionfreetextbasedediting}. 
\method adopts this inversion-free formulation for audio editing, while \methodplus further introduces optimal-transport coupling to regularize the source-target trajectory, reducing ambiguity while preserving prompt-driven semantic changes.

\section{Background}

\begin{figure*}[t]
    \centering
    \includegraphics[width=0.99\textwidth]{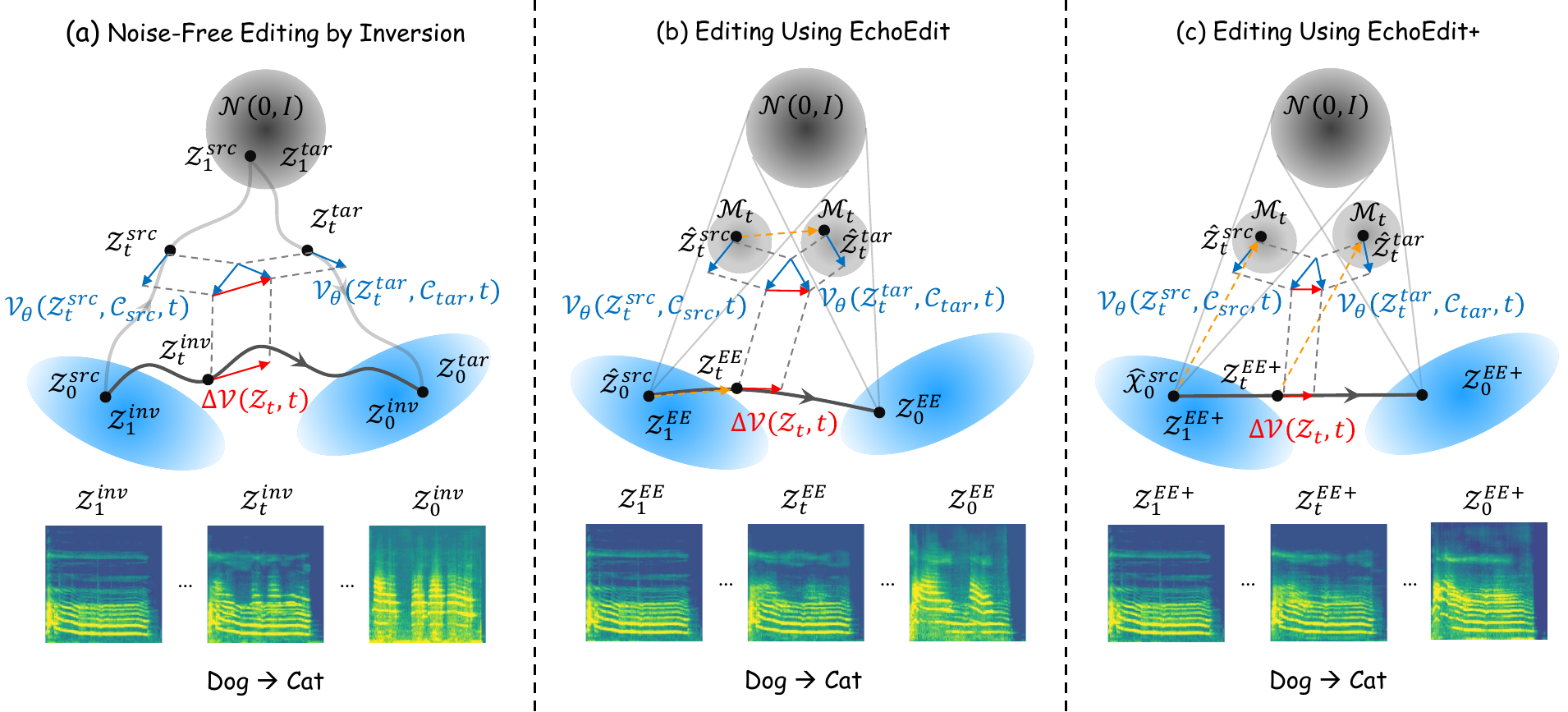}
    \caption{\textbf{Comparison of editing field coupling geometries.}
    At noise level $t$, the current edited latent $\vect{y}$ induces the noisy marginal $\mathcal{M}_t(\vect{y})=\mathcal{L}((1-t)\vect{y}+t\vect{\epsilon})$.
    (a) Inversion/noising methods edit through a source-to-noise-to-target bridge; the noisy intermediate state exposes generative freedom but weakens source structure.
    (b) \method constructs a direct residual field by contrasting source- and target-conditioned velocities, avoiding source inversion, but its translated stochastic target probe need not lie on $\mathcal{M}_t(\vect{y})$.
    (c) \methodplus retains the direct residual while imposing the OT-admissible coupling $\pi_t^{+}\in\Pi(\mathcal{M}_t(\vect{x}^{\src}),\mathcal{M}_t(\vect{y}))$.
    This minimum-cost coupling keeps model queries on the appropriate rectified-flow audio manifold, stabilizing the editing trajectory and reducing uncertainty drift.}
    \label{fig:schematic}
\end{figure*}

\subsection{Conditional Audio Flow and the Editing Problem}
\label{sec:prelim_audio_flow}

We work in the latent space of a pretrained audio generator.
Let $\vect{x}\in\mathbb{R}^{d}$ denote a clean audio latent, let $\vect{\epsilon}\sim\mathcal{N}(0,I)$ denote a Gaussian endpoint, and let $c$ be a text condition.
A rectified-flow audio model defines a conditional velocity field through the probability-flow ODE
\begin{equation}
    \label{eq:audio_rf_ode}
    \frac{d\vect{z}_t}{dt}
    =
    V_\theta(\vect{z}_t,t,c),
    \qquad t\in[0,1],
\end{equation}
where $t=0$ corresponds to the clean audio endpoint and $t=1$ to the Gaussian endpoint.
The rectified-flow training path is the linear interpolation~\cite{liu2022flowstraightfastlearning,lipman2023flowmatchinggenerativemodeling,albergo2023buildingnormalizingflowsstochastic}
\begin{equation}
    \label{eq:audio_rf_interpolant}
    \vect{z}_t=(1-t)\vect{x}+t\vect{\epsilon}.
\end{equation}
Sampling integrates \eqref{eq:audio_rf_ode} from high noise to the clean endpoint.
Audio models such as Stable Audio 3 instantiate this flow in a learned latent space rather than directly on waveforms, but the transport view remains unchanged~\cite{evans2026stableaudio3,parker2026samesemanticallyalignedmusicautoencoder}.

Text control is commonly strengthened by classifier-free guidance~\cite{ho2022classifierfreediffusionguidance}.
We write the guided velocity as
\begin{equation}
    \label{eq:audio_cfg_velocity}
    V_\theta^{w}(\vect{z},t,c)
    =
    V_\theta(\vect{z},t,\varnothing)
    +
    w\left[
    V_\theta(\vect{z},t,c)-V_\theta(\vect{z},t,\varnothing)
    \right].
\end{equation}
Given a source prompt $c_{\src}$, a target prompt $c_{\tar}$, and a source latent $\vect{x}^{\src}\sim p_0(\cdot\mid c_{\src})$, editing aims to obtain a latent $\vect{x}^{\tar}$ that is compatible with $p_0(\cdot\mid c_{\tar})$ while preserving source attributes not involved in the prompt modification.
Unlike ordinary conditional generation, the target latent should approach the target-prompt manifold while remaining anchored to the source latent in terms of timing, texture, and structural characteristics.

\subsection{Observable Velocity Differences at Noisy States}
\label{sec:observable_velocity}

The pretrained model exposes conditional velocities only through queries at noisy states.
For a clean anchor $\vect{x}$, define the rectified-flow corruption kernel
\begin{equation}
    \label{eq:audio_kernel}
    K_t(\cdot\mid \vect{x})
    =
    \mathcal{L}\big((1-t)\vect{x}+t\vect{\epsilon}\big),
    \qquad \vect{\epsilon}\sim\mathcal{N}(0,I),
\end{equation}
where $\mathcal{L}(\cdot)$ denotes the law of a random variable.
Let $V^{\src}(\vect{z},t)=V_\theta^{w_{\src}}(\vect{z},t,c_{\src})$ and $V^{\tar}(\vect{z},t)=V_\theta^{w_{\tar}}(\vect{z},t,c_{\tar})$.
For a current edited latent $\vect{y}$, consider a coupling $\pi_t(\cdot,\cdot\mid \vect{x}^{\src},\vect{y})$ between a source noisy state and a target-side noisy state.
The observable editing field induced by this coupling is
\begin{equation}
    \label{eq:observable_residual}
    \begin{aligned}
    \mathcal{R}_t(\vect{y};\vect{x}^{\src},\pi)
    =
    \mathbb{E}_{\pi_t}\big[
    V^{\tar}(\vect{z}^{\tar},t)
    -
    V^{\src}(\vect{z}^{\src},t)
    \big].
    \end{aligned}
\end{equation}
This expression is the common observable object behind direct velocity-difference editing.
It compares how the same pretrained flow moves under the source and target conditions, so shared generative motion can cancel while the residual carries the intended edit.
The quality of the field, however, depends on the coupling $\pi_t$.
If the target-side noisy state is shifted away from the rectified-flow marginal associated with the current edited latent, the model is queried off its training manifold; the resulting residual may be biased or high variance.

\begin{algorithm}[t]
    \caption{\textbf{Simplified algorithm for \methodplus}}
    \label{alg:EchoEditPlus}
    \textbf{Input}: source latent $X^{\mathrm{src}}$, prompts $c_{\mathrm{src}},c_{\mathrm{tar}}$, schedule $\{t_i\}_{i=0}^{T}$, starting index $\nmax$.\\
    \textbf{Output}: edited latent $X^{\mathrm{tar}}$.
    \begin{algorithmic}[1]
    \State $Z^{+}_{t_{\nmax}} \leftarrow X^{\mathrm{src}}$
    \For{$i=\nmax$ \textbf{to} $1$}
        \State Sample $\epsilon_i\sim\mathcal{N}(0,I)$
        \State $\hat{Z}^{\src}_{t_i}\leftarrow (1-t_i)X^{\mathrm{src}}+t_i\epsilon_i$
        \State $\hat{Z}^{\tar,+}_{t_i}\leftarrow (1-t_i)Z^{+}_{t_i}+t_i\epsilon_i$
        \State $V^{\Delta}_{t_i}\leftarrow
        V^{\tar}(\hat{Z}^{\tar,+}_{t_i},t_i)
        -
        V^{\src}(\hat{Z}^{\src}_{t_i},t_i)$
        \State $Z^{+}_{t_{i-1}}\leftarrow Z^{+}_{t_i}+(t_{i-1}-t_i)V^{\Delta}_{t_i}$
    \EndFor
    \State \textbf{return} $X^{\mathrm{tar}}=Z^{+}_{0}$
    \end{algorithmic}
\end{algorithm}

\subsection{Inversion-Based Editing as a Reference Topology}
\label{sec:inversion_audio}

Inversion-based editors choose the coupling indirectly through a noisy intermediate state.
Starting from $\vect{z}^{\src}_0=\vect{x}^{\src}$, the source-conditioned trajectory is moved toward noise by
\begin{equation}
    \label{eq:audio_inv_forward}
    \frac{d\vect{z}^{\src}_t}{dt}
    =
    V^{\src}(\vect{z}^{\src}_t,t),
\end{equation}
until an edit depth $t_m$ is reached.
The target-conditioned trajectory is then initialized at the same noisy state and integrated back toward the clean endpoint:
\begin{equation}
    \label{eq:audio_inv_reverse}
    \frac{d\vect{z}^{\tar}_t}{dt}
    =
    V^{\tar}(\vect{z}^{\tar}_t,t),
    \qquad
    \vect{z}^{\tar}_{t_m}=\vect{z}^{\src}_{t_m}.
\end{equation}
For an increasing schedule $\{t_i\}_{i=0}^{T}\subset[0,1]$, the source pass and target pass are discretized as
\begin{equation}
    \label{eq:audio_inv_forward_discrete}
    \vect{z}_{t_i}^{\src}
    =
    \vect{z}_{t_{i-1}}^{\src}
    +(t_i-t_{i-1})
    V^{\src}(\vect{z}_{t_{i-1}}^{\src},t_{i-1}),
\end{equation}
and
\begin{equation}
    \label{eq:audio_inv_reverse_discrete}
    \vect{z}_{t_{i-1}}^{\tar}
    =
    \vect{z}_{t_i}^{\tar}
    +(t_{i-1}-t_i)
    V^{\tar}(\vect{z}_{t_i}^{\tar},t_i).
\end{equation}
SDEdit follows the same reverse target sampler but replaces the source ODE with a direct draw from the corruption kernel~\cite{meng2022sdeditguidedimagesynthesis}:
\begin{equation}
    \label{eq:audio_sdedit_noise}
    \tilde{\vect{z}}_{t_m}^{\src}
    =
    (1-t_m)\vect{x}^{\src}+t_m\vect{\epsilon}.
\end{equation}

ODE inversion and SDEdit differ in how the noisy edit state is obtained, but both impose a source-to-noise-to-target topology.
At large $t_m$, the noisy state weakens onset phase, transient sharpness, groove, room texture, and other cues that determine whether the edited audio still sounds like the source.
At small $t_m$, the target sampler remains close to the source but has limited freedom to satisfy the new prompt.

\section{Method}

As shown in Figure~\ref{fig:schematic}, our goal is to replace the inversion bridge with a direct source-to-target audio transport that remains aligned with the rectified-flow noisy manifolds throughout editing.
The key challenge is that a direct velocity residual is observable from the pretrained audio flow, but the coupling used to compare source- and target-conditioned noisy states is not determined by the model itself.
\method first removes the source-to-noise-to-target detour by estimating this residual through stochastic source marginals.
\methodplus then resolves the remaining geometric ambiguity by imposing an OT-admissible coupling, so that the target-side velocity query remains compatible with the current edited latent rather than drifting away from the audio-flow manifold.

\subsection{Editing as Coupled Transport}

The inversion topology can be viewed as one particular coupling between a source-conditioned trajectory and a target-conditioned trajectory.
We instead formulate editing directly as a coupled transport problem.
Let $\vect{y}$ denote the current clean-domain edited latent, initialized at $\vect{x}^{\src}$.
At flow noise level $t$, the source branch is distributed according to $K_t(\cdot\mid \vect{x}^{\src})$, while an admissible target branch should remain compatible with the rectified-flow marginal $K_t(\cdot\mid \vect{y})$.
Thus, the local editing direction is determined by a coupling
\begin{equation}
    \pi_t\in
    \Pi\!\left(
    K_t(\cdot\mid \vect{x}^{\src}),
    K_t(\cdot\mid \vect{y})
    \right),
\end{equation}
where $\Pi(\mu,\nu)$ denotes the set of joint distributions with marginals $\mu$ and $\nu$.
Given a coupling, the observable residual field is the coupled velocity difference defined in \eqref{eq:observable_residual}.

This view separates two questions that are entangled in inversion methods.
First, which residual should move the source toward the target prompt?
Second, which source-target noisy pairing should be used to measure that residual?
The first question is answered by the pretrained conditional flow.
The second is a transport problem.
Following the dynamic optimal-transport perspective~\cite{benamou2000computational,peyre2019computational}, an editing field should avoid unnecessary motion while satisfying the target condition.
In our setting, this principle is imposed locally at every flow noise level by selecting a low-cost coupling before measuring the velocity difference.

\begin{figure*}[t]
    \centering
    \includegraphics[width=\textwidth]{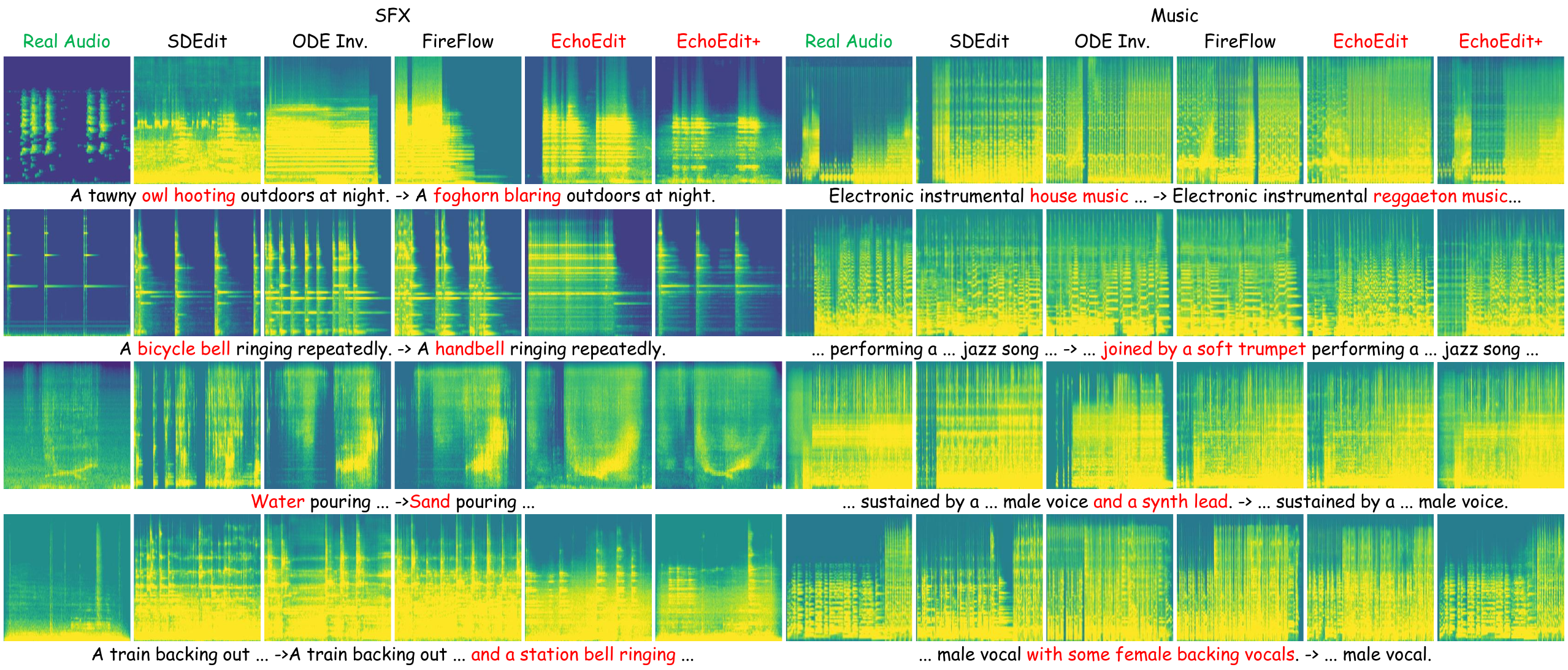}
    \caption{\textbf{Qualitative comparison of edited audio.}
    In both sound-effect (left) and music (right) editing, \methodplus achieves better target-prompt adherence while preserving the structural characteristics of the original audio.}
    \label{fig:qualitative_audio}
\end{figure*}

\subsection{\method as a Stochastic Estimator}

\method is the direct, inversion-free estimator obtained by replacing the deterministic inversion trajectory with stochastic source marginals.
It samples noisy source states from $K_t(\cdot\mid\vect{x}^{\src})$ and compares source- and target-conditioned velocities under a shared stochastic perturbation.
Equivalently, \method uses a translation coupling $\pi_t^{\mathrm{A}}$ whose source-target displacement is the current clean-domain edit displacement and is independent of the noise level $t$.
The corresponding editing field is
\begin{equation}
    \label{eq:audioedit_field}
    u_t^{\mathrm{A}}(\vect{y})
    =
    \mathcal{R}_t(\vect{y};\vect{x}^{\src},\pi_t^{\mathrm{A}}).
\end{equation}
Integrating this field yields a direct source-to-target editing trajectory without source inversion, latent optimization, paired edit data, or attention-map intervention.

At high noise levels the residual tends to capture coarse semantic changes, such as sound class, instrument family, texture, or style.
At low noise levels it has access to finer acoustic details, including transient envelopes, room color, and microtiming.
This gives \method a coarse-to-fine interpretation while avoiding the explicit source-to-noise-to-target detour of inversion.

\subsection{Uncertainty Drift and Manifold Deviation}

The same stochasticity that makes \method inversion-free also introduces a structural limitation.
The translation coupling $\pi_t^{\mathrm{A}}$ preserves a full clean-domain displacement at every noise level.
However, the rectified-flow marginal associated with the current edited latent contracts this displacement as $t$ increases.
Consequently, the target-side marginal induced by $\pi_t^{\mathrm{A}}$ does not generally equal $K_t(\cdot\mid\vect{y})$.
For the linear rectified-flow kernel in \eqref{eq:audio_kernel}, the mismatch in the target-side mean is
\begin{equation}
    \label{eq:audioedit_mismatch}
    m_t^{\mathrm{A}}(\vect{y})-(1-t)\vect{y}
    =
    t(\vect{y}-\vect{x}^{\src}),
\end{equation}
where $m_t^{\mathrm{A}}(\vect{y})$ denotes the mean of the target-side noisy variable under the \method coupling.
Thus the off-manifold displacement grows with both the noise level and the accumulated edit magnitude.

We refer to this phenomenon as uncertainty drift.
It is not simply a consequence of Monte Carlo variance, but rather a marginal mismatch induced by the coupling mechanism itself.
As the edited latent deviates from the source trajectory, the target velocity may be evaluated at noisy states that gradually depart from the rectified-flow training manifold of $\vect{y}$.
In audio editing, such deviations can undermine structural consistency even when the expected semantic residual remains informative.
These small latent perturbations may manifest as transient smearing, rhythmic drift, phase instability, or degraded musical continuity.

\subsection{OT-Regularized \methodplus}

\methodplus resolves the mismatch by choosing the coupling through an optimal-transport criterion.
At each flow noise level, we select the quadratic-cost coupling between the source noisy marginal and the edited noisy marginal:
\begin{equation}
    \label{eq:ot_coupling}
    \begin{aligned}
    \pi_t^{+}
    =
    \arg\min_{\pi\in
    \Pi\left(
    K_t(\cdot\mid \vect{x}^{\src}),
    K_t(\cdot\mid \vect{y})
    \right)}
    \mathbb{E}_{\pi}\big[
    \|\vect{z}^{\tar}-\vect{z}^{\src}\|_2^2
    \big].
    \end{aligned}
\end{equation}
For the rectified-flow kernel, the two marginals in \eqref{eq:ot_coupling} are translations of the same isotropic Gaussian.
The optimal Monge coupling therefore aligns the same Gaussian perturbation in both branches:
\begin{equation}
    \label{eq:ot_monge}
    \vect{z}^{\src}_t=(1-t)\vect{x}^{\src}+t\vect{\epsilon},
    \qquad
    \vect{z}^{\tar,+}_t=(1-t)\vect{y}+t\vect{\epsilon}.
\end{equation}
This is the unique quadratic-cost coupling up to null sets for the Gaussian base measure.
It ensures that the target-side noisy variable has the rectified-flow marginal $K_t(\cdot\mid\vect{y})$.

The resulting \methodplus editing field is
\begin{equation}
    \label{eq:audioeditplus_field}
    u_t^{+}(\vect{y})
    =
    \mathcal{R}_t(\vect{y};\vect{x}^{\src},\pi_t^{+}).
\end{equation}
Compared with the \method coupling, the OT coupling reduces the instantaneous paired displacement from a full clean-domain displacement to the contracted displacement required by the rectified-flow geometry.
Under the quadratic cost,
\begin{equation}
    \label{eq:ot_contraction}
    \mathbb{E}_{\pi_t^{+}}\|\vect{z}^{\tar}-\vect{z}^{\src}\|_2^2
    =
    (1-t)^2\|\vect{y}-\vect{x}^{\src}\|_2^2
    \le
    \|\vect{y}-\vect{x}^{\src}\|_2^2.
\end{equation}
Thus, as the trajectory approaches high-noise states, the coupled branches converge rather than carrying an unnecessary displacement through the noisy manifold.
At the clean endpoint, the full edit displacement is recovered.
This behavior gives \methodplus the desired closed loop: it performs semantic transport toward the target prompt while keeping all velocity queries on the appropriate audio-flow manifold.
Algorithm~\ref{alg:EchoEditPlus} gives the simplified procedure, the full version is provided in the supplementary material.

\subsection{Transport-Cost Diagnostic Against Inversion}
\label{sec:exact_inversion_audio}

To isolate the geometric properties of the edit path, we construct a synthetic transport diagnostic using model-generated audio samples.
The source clips are generated by Stable Audio 3 from source prompts with the same backbone used by the editor, ensuring that they reside close to the model's own generative manifold.
This controlled setting eliminates major confounding factors in real recordings, such as compression artifacts, background noise, and out-of-distribution reconstruction errors.
Sound-effect prompts are generated from controlled acoustic templates, and target prompts are constructed by applying minimal concept-level modifications, including substitution, addition, or removal of acoustic attributes while preserving timing descriptors.
Music prompts are derived from Song Describer Dataset captions and modified through minimal keyword-level edits, including replacing, adding, or removing concepts such as instruments or genres while keeping the remaining caption content fixed.

\methodplus achieves the lowest latent and acoustic transport cost on both domains.
Compared with the original stochastic \method coupling, \methodplus reduces latent MSE from 0.41 to 0.10 on sound effects and from 0.64 to 0.12 on music.
To directly verify uncertainty drift, we measure the trajectory-wise Wasserstein marginal mismatch between the target query used by \method and the admissible rectified-flow marginal.
The mismatch is nonzero for \method (mean/max $W_2^{\mathrm{MM}}$: $0.11/0.15$ for sound effects and $0.14/0.19$ for music), correlates with degraded preservation, and is eliminated by the OT coupling in \methodplus (mean/max $0.00/0.00$ in both domains).
These results indicate that OT-regularized direct transport reduces unnecessary latent displacement while remaining target-aligned.

\begin{table}[t]
    \centering
    \small
    \begin{tabular}{@{}lccc@{}}
        \toprule
        \textbf{Method} & \textbf{MOS-T} $\uparrow$ & \textbf{MOS-P} $\uparrow$ & \textbf{Overall} $\uparrow$ \\
        \midrule
        FireFlow & 2.822 & 2.676 & 2.749 \\
        ODE Inv. & 2.958 & 2.836 & 2.897 \\
        SDEdit & 3.276 & 3.109 & 3.192 \\
        \method & 4.358 & 4.467 & 4.412 \\
        \textbf{\methodplus} & \textbf{4.393} & \textbf{4.580} & \textbf{4.487} \\
        \bottomrule
    \end{tabular}
    \caption{Subjective evaluation on a 5-point Likert scale.}
    \label{tab:subjective}
\end{table}

\section{Experiments}
\label{sec:experimental_setup}

\subsection{Experimental Setup}

\begin{table*}[t]
    \centering
    \small
    \begin{tabular}{llccccccc}
        \toprule
        \textbf{Domain} & \textbf{Method} & \textbf{CLAP-T} $\uparrow$ & \textbf{CLAP-A} $\uparrow$ & \textbf{LSD} $\downarrow$ & \textbf{MCD} $\downarrow$ & \textbf{LPAPS} $\downarrow$ & \textbf{Structure} $\uparrow$ & \textbf{FAD} $\downarrow$ \\
        \midrule
        \multirow{5}{*}{SFX} & FireFlow & 0.36 & 0.36 & 30.09 & 685.69 & 0.32 & 0.45 & 83.36 \\
        & ODE Inv. & 0.41 & 0.40 & 28.92 & 688.37 & 0.29 & 0.48 & 76.65 \\
        & SDEdit & 0.44 & 0.46 & 26.20 & 651.16 & 0.28 & 0.48 & 76.24 \\
        & \method & \textbf{0.53} & 0.59 & 23.34 & 575.80 & 0.22 & 0.56 & 62.36 \\
        & \textbf{\methodplus} & \textbf{0.53} & \textbf{0.68} & \textbf{19.01} & \textbf{499.73} & \textbf{0.17} & \textbf{0.59} & \textbf{50.82} \\
        \midrule
        \multirow{5}{*}{Music} & FireFlow & 0.57 & 0.66 & 24.02 & 610.39 & 0.25 & 0.89 & 51.41 \\
        & ODE Inv. & 0.59 & 0.68 & 23.15 & 607.08 & 0.24 & 0.90 & 49.38 \\
        & SDEdit & 0.56 & 0.67 & 20.03 & 542.97 & 0.23 & 0.91 & 53.97 \\
        & \method & \textbf{0.61} & 0.71 & 20.26 & 497.57 & 0.20 & 0.92 & 46.90 \\
        & \textbf{\methodplus} & \textbf{0.61} & \textbf{0.80} & \textbf{16.38} & \textbf{350.10} & \textbf{0.11} & \textbf{0.98} & \textbf{25.01} \\
        \bottomrule
    \end{tabular}
    \caption{\textbf{Comparison on real sound-effect and music edits.}
    The best results in each numeric column are highlighted in bold.
    Full operation-type and sensitivity breakdowns are provided in the supplementary material.}
    \label{tab:main}
\end{table*}

\paragraph{Benchmarks and protocol.}
We evaluate on two real-audio editing benchmarks designed to stress source preservation under prompt changes.
The sound-effect split contains 222 edits derived from FSD50K~\cite{Fonseca_2022}, spanning environmental events, object interactions, percussive sounds, and short acoustic textures.
The music split contains 198 edits derived from the Song Describer Dataset~\cite{manco2023songdescriberdatasetcorpus}, using 30-second excerpts whose prompts alter instrument, genre, production character, or ensemble description while preserving the surrounding musical context.
Each instance consists of a source waveform, a source prompt, and a target prompt.
The main benchmark covers replacement, insertion, and deletion edits; style transfer, source-prompt sensitivity, chained editing, and duration scaling are reported in the supplementary material.

\paragraph{Baselines and fairness.}
All methods use the same Stable Audio 3 medium backbone, SAME autoencoder, 44.1 kHz decoding, prompt pairs, durations, and random seeds whenever applicable.
This isolates the editing mechanism from differences in generator architecture, codec, or training data.
We compare against representative inversion/noising baselines on the same backbone: SDEdit~\cite{meng2022sdeditguidedimagesynthesis}, SA3 ODE inversion, and FireFlow~\cite{deng2024fireflowfastinversionrectified}.
SDEdit edits by corrupting the source and denoising under the target prompt; ODE inversion and FireFlow both follow a two-pass source-to-noise-to-target topology, with FireFlow using a midpoint solver.
We do not use TangoFlux~\cite{hung2024tangoflux} or FluxMusic~\cite{fei2024fluxmusic} as alternative backbones, as our preliminary evaluation found that their current generation capabilities did not provide reliable source-conditioned audio-to-audio editing under the required prompt-pair setting.
We likewise do not include audio adaptations of iRFDS~\cite{yang2025irfds} or RF-Inversion~\cite{rout2024semanticimageinversionediting} as main baselines, as they did not achieve competitive audio editing performance in our setting.

\paragraph{Metrics.}
We evaluate target alignment, source preservation, and distributional realism.
CLAP-T measures target-text alignment, while CLAP-A measures semantic preservation of the source audio~\cite{wu2024largescalecontrastivelanguageaudiopretraining}.
LSD, MCD, and LPAPS quantify spectral, cepstral, and perceptual-feature distortion~\cite{SpecVQGAN_Iashin_2021,paissan2023audio,Kong_2020}.
For structure, we use AudioBERTScore on sound effects and MelodySim on music~\cite{Kishi2026AudioBERTScore,lu2025melodysim}; FAD measures set-level realism~\cite{Kilgour_2019}.
We further conduct a 5-point MOS study on anonymized edited outputs, with separate ratings for target alignment and source preservation.
Overall MOS is computed as the mean of the two perceptual axes.

\subsection{Comparison with Prior Methods}

\paragraph{Subjective results.}
The spectrogram examples in Figure~\ref{fig:qualitative_audio} show comparisons between \methodplus and the competing methods. Inversion baselines often introduce broad spectral reorganization, whereas \methodplus better preserves event layout and background structure while changing the requested acoustic concept.
Automatic audio metrics only partially capture edit naturalness, especially when small phase or transient changes become perceptually salient.
We therefore complement the objective evaluation with a MOS study, reported in Table~\ref{tab:subjective}.
Based on 2,250 scored method instances from 10 participants, \methodplus achieves the highest target-alignment, preservation, and overall MOS scores.
\begin{figure}[t]
    \centering
    \includegraphics[width=\columnwidth]{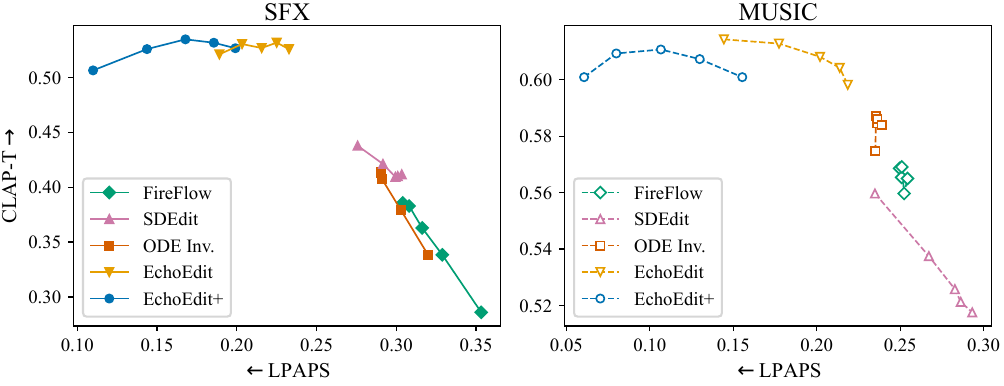}
    \caption{\textbf{Target-alignment/source-preservation frontier.}
    CLAP-T versus LPAPS under guidance and strength sweeps for sound-effect and music edits; the desired region is upper-left.}
    \label{fig:pareto}
\end{figure}
Relative to \method, the OT-stabilized editor yields a modest gain in target alignment (+0.035) and a larger gain in preservation (+0.113), producing a +0.075 improvement in overall MOS.
Compared with the strongest inversion/noising baseline, SDEdit, \methodplus improves MOS-T by 1.117 and MOS-P by 1.471.
This pattern supports the proposed view that OT regularization primarily improves perceived source consistency and naturalness while retaining the semantic effectiveness of inversion-free direct transport.
The full protocol and operation-level analysis are provided in the supplementary material.

\begin{figure}[t]
    \centering
    \begin{minipage}[t]{0.43\linewidth}
      \centering
      \includegraphics[width=\linewidth]{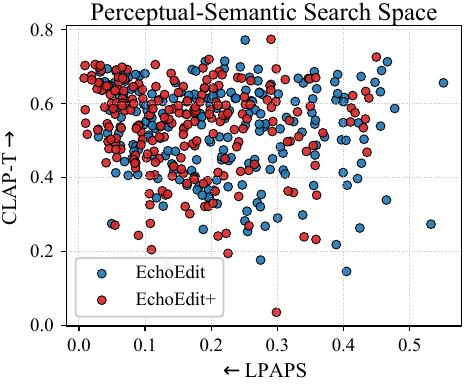}    
    \end{minipage}
    \begin{minipage}[t]{0.43\linewidth}
      \centering
      \includegraphics[width=\linewidth]{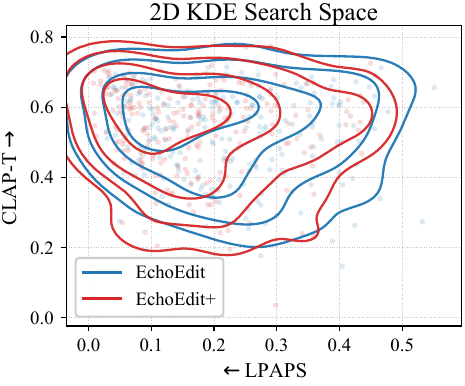}
    \end{minipage}
      \begin{minipage}[t]{0.43\linewidth}
      \centering
      \includegraphics[width=\linewidth]{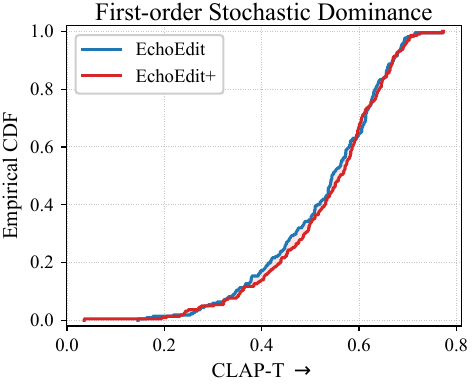}    
    \end{minipage}
    \begin{minipage}[t]{0.43\linewidth}
      \centering
      \includegraphics[width=\linewidth]{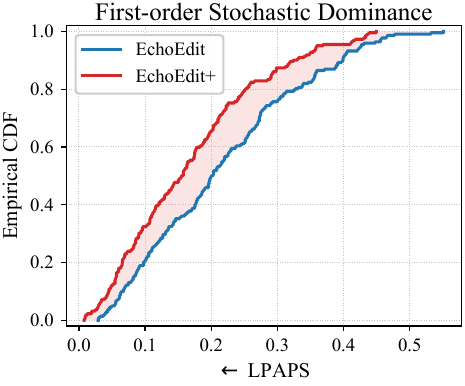}
    \end{minipage}
  \caption{\textbf{Search-space and distributional evidence for OT stabilization.}
  On representative sound-effect edits, the upper panels show the CLAP-T/LPAPS search space and its density contours, while the lower panels summarize the empirical distributions.
  \methodplus preserves the target-alignment regime of \method while shifting mass toward lower LPAPS, indicating that OT regularization tightens source consistency without weakening semantic transport.}
  \label{fig:ccf_effort}
  \end{figure}

\paragraph{Objective results.}

The real-audio comparison in Table~\ref{tab:main} reveals a consistent semantic-preservation pattern across sound effects and music.
Relative to inversion/noising baselines, \method improves target alignment while reducing spectral, cepstral, and perceptual distortion, indicating that direct conditional transport avoids much of the source-information loss induced by the source-to-noise-to-target route.
\methodplus improves the preservation side of this direct editor while retaining essentially the same target alignment.
This is the empirical signature predicted by our theory, where OT regularization preserves source structure without sacrificing target alignment by reducing coupling-induced uncertainty drift and keeping velocity queries closer to the rectified-flow audio manifold.

Figure~\ref{fig:pareto} presents the editing results of all methods for sound effects and music under varied hyperparameter settings.
Direct editors occupy a stronger alignment-preservation frontier than inversion baselines, and \methodplus shifts the frontier toward lower perceptual distortion, consistent with OT acting as a geometric stabilizer.
The distributional analysis in Figure~\ref{fig:ccf_effort} (top) further isolates the mechanism.
\methodplus remains in the high-CLAP-T region of \method, while raw samples and density contours move toward lower LPAPS.
The gain therefore arises from suppressing noisy-marginal drift and improving source consistency rather than attenuating the target edit, enabling our method to strictly first-order stochastically dominate competing approaches(Figure~\ref{fig:ccf_effort}, bottom).
Complete visualizations and quantitative evaluation are provided in the supplementary material.

\section{Conclusion and limitations}

We presented \method and \methodplus for zero-shot text-guided audio editing with pretrained rectified-flow models.
\method reformulates editing as direct source-to-target transport by replacing inversion with stochastic velocity-difference estimation, while \methodplus further stabilizes this transport through an optimal-transport coupling that preserves consistency with the rectified-flow audio manifold.
Our theoretical analysis reveals uncertainty drift as a coupling-induced marginal mismatch that can drive edited trajectories away from the source-conditioned manifold.
Experiments demonstrate that OT stabilization effectively reduces such drift, improving source preservation, structural consistency, and perceptual quality while maintaining target-prompt alignment.
These results establish direct OT-regularized transport as a practical and training-free paradigm for controllable audio editing without source inversion or edit-specific supervision.
However, \methodplus is primarily designed for controlled, source-preserving edits, and its strong preservation capability can become restrictive when substantial semantic changes are required, such as replacing acoustically incompatible events, modifying multiple musical roles simultaneously, or transforming vocal and arrangement structures. 
In such cases, greater editing flexibility can be achieved by simply removing the source-audio dependency in the final generation step.

\bibliography{aaai2027}

\newpage
\appendix
\setcounter{secnumdepth}{1}

\section{Full Related Work}
\label{app:full_related_work}

\paragraph{Text-conditioned audio generation.}
Text-guided audio editing is built on progress in large-scale text-conditioned generation.
Autoregressive token models such as AudioGen and MusicGen learn temporal structure through discrete audio codes and language conditioning, enabling controllable generation of environmental sounds and music~\cite{kreuk2023audiogentextuallyguidedaudio,copet2024simplecontrollablemusicgeneration}.
Latent diffusion and flow-based systems, including AudioLDM, AudioLDM 2, Tango, Make-An-Audio, and Stable Audio, instead synthesize in compressed continuous representations and decode the resulting latents to waveform space~\cite{liu2023audioldmtexttoaudiogenerationlatent,liu2023audioldm2,ghosal2023texttoaudiogenerationusinginstructiontuned,huang2023makeanaudio,evans2026stableaudio3}.
More recent rectified-flow audio generators such as TangoFlux and FluxMusic further advance this paradigm by combining flow-matching objectives with transformer backbones, enabling faster sampling and stronger text-audio alignment~\cite{hung2024tangoflux,fei2024fluxmusic}.
The Stable Audio 3 family is especially relevant to this work because it combines a rectified-flow transformer with the SAME semantically aligned audio autoencoder, producing a continuous latent geometry in which probability-flow trajectories can be interpreted as transport paths~\cite{evans2026stableaudio3,parker2026samesemanticallyalignedmusicautoencoder}.
These generators provide the semantic and acoustic prior reused by editors, but generation alone does not define the source-conditioned editing problem.
An editor must satisfy a new text prompt while retaining source-specific information that may be absent from the prompt, such as timing, attack structure, room coloration, rhythmic phase, melodic contour, and arrangement identity.

\paragraph{Instruction-trained and task-supervised audio editing.}
A second line of work trains editing behavior directly.
AUDIT, InstructME, EditGen, and subsequent audio-language editing frameworks broaden the interface from fixed editing labels to natural-language instructions, covering general audio, music, and increasingly free-form multi-type operations~\cite{wang2023auditaudioeditingfollowing,han2023instructmeinstructionguidedmusic,sioros2025editgenharnessingcrossattentioncontrol,ungersbock2025saoinstructfreeformaudioediting,lan2025guidingaudioeditingaudio,tao2026mmeditunifiedframeworkmultitype}.
This paradigm can learn rich correspondences between instructions and output changes, and can incorporate operation-specific priors unavailable to a generic generator.
Its limitation is that the resulting editor is coupled to the coverage, annotation quality, and distributional assumptions of the edit corpus.
It also changes the methodological question: performance is partly a property of supervised edit data and task-specific adaptation.
The setting studied here is deliberately different.
\method and \methodplus are training-free, use a pretrained text-conditioned audio flow at inference time, and do not require paired edit examples, temporal masks, operation labels, or model finetuning.
This motivates a derivation in terms of observable velocity fields rather than a learned edit operator.

\paragraph{Zero-shot editing by noising, inversion, and optimization.}
Most zero-shot audio editors inherit the topology developed for diffusion-based image editing: move the source toward a noisy latent state, then regenerate under the target condition.
SDEdit realizes this principle by corrupting the source to an intermediate noise level and denoising it under the target prompt~\cite{meng2022sdeditguidedimagesynthesis}.
DDPM-inversion audio editing and MusicMagus reconstruct a source trajectory before changing the conditioning signal, while inference-time optimization methods such as DITTO optimize latent or guidance variables to improve consistency, especially for music~\cite{manor2024zeroshotunsupervisedtextbasedaudio,zhang2024musicmaguszeroshottexttomusicediting,novack2024dittodiffusioninferencetimetoptimization}.
Prompt-guided precise audio editing and attention-control methods add localization or structural constraints, but still typically rely on a corrupted trajectory, model-internal control, or test-time adjustment~\cite{xu2024promptguidedpreciseaudioediting,sioros2025editgenharnessingcrossattentioncontrol}.
These approaches are effective because the noisy bridge gives the target prompt generative freedom.
However, the same bridge creates the preservation--editability trade-off: shallow corruption preserves the source but may under-edit, whereas deep corruption may alter event timing, transient envelopes, background texture, phase coherence, or musical identity.
This trade-off is more severe in audio than in many visual settings because small temporal or phase deviations are perceptually salient.

\paragraph{Flow-based editing and inversion acceleration.}
Rectified flow and flow matching replace a long denoising chain with a continuous velocity field that transports between noise and data distributions through a probability-flow ODE~\cite{liu2022flowstraightfastlearning,lipman2023flowmatchinggenerativemodeling,albergo2023buildingnormalizingflowsstochastic}.
This formulation changes the editing question.
When a model exposes conditional velocities, one can ask whether editing must still reconstruct a source-to-noise-to-target path, or whether source and target conditions can be compared directly through the velocity field.
In computer vision, recent advances have demonstrated that inversion-free direct velocity differences can achieve source-preserving edits with pretrained flow models~\cite{kulikov2025floweditinversionfreetextbasedediting}.
Other flow-editing works improve or reinterpret inversion, including FireFlow, rectified-flow inversion methods, and rectified-flow priors for downstream editing tasks, which seek more accurate or efficient two-pass trajectories~\cite{deng2024fireflowfastinversionrectified,rout2024semanticimageinversionediting,wang2025tamingrectifiedflowinversion,yang2025irfds}.
In audio, RFM-Editing studies rectified-flow matching for text-guided editing, showing that flow-based audio backbones can support edit operations beyond unconditional generation~\cite{gao2026rfmeditingrectifiedflowmatching}.
These works motivate velocity fields as the primitive object for editing.
They do not, however, resolve the geometric question that arises in a stochastic direct estimator: how should the source-side and target-side noisy variables be coupled when the current edited latent is no longer equal to the source?

\paragraph{Optimal transport and flow geometry.}
Optimal transport provides a variational language for precisely this coupling question.
The Benamou--Brenier formulation characterizes transport dynamically through a density path and a velocity field whose kinetic energy is minimized subject to the continuity equation~\cite{benamou2000computational,peyre2019computational}.
Flow matching and rectified flow are naturally compatible with this view because their interpolants define probability paths and velocity fields between simple and data distributions~\cite{liu2022flowstraightfastlearning,lipman2023flowmatchinggenerativemodeling}.
For source-preserving editing, however, the relevant OT object is not only a global map from a source distribution to a target distribution.
The practical estimator must evaluate source- and target-conditioned velocities at paired noisy states, so the local coupling between noisy marginals determines whether the residual is measured on the correct audio-flow manifold.
\methodplus uses OT at this local coupling level.
The semantic direction still comes from conditional velocity differences, while the OT criterion selects the lowest-cost admissible pairing between the source noisy marginal and the noisy marginal of the evolving edited latent.
This separation between semantic transport and coupling geometry is the main distinction from both inversion-based methods and unregularized direct residual fields.

\paragraph{Position of \method and \methodplus.}
\method occupies the training-free, inversion-free region of the audio editing design space.
It does not reconstruct the source into a noise latent, optimize a latent trajectory, or rely on model-internal attention maps.
Instead, it estimates a direct editing field by contrasting source- and target-conditioned velocities under a shared stochastic probe.
This construction removes the inversion bottleneck and gives a natural coarse-to-fine editing interpretation, but it leaves the coupling underdetermined.
When the translated target-side stochastic variable does not have the rectified-flow marginal of the current edited latent, the residual can be evaluated off manifold; this is the uncertainty drift analyzed in the main paper.
\methodplus resolves this missing degree of freedom by selecting the OT-admissible coupling with minimal quadratic cost.
Consequently, the paper contributes both an inversion-free audio editing framework and a geometric explanation of why OT stabilization improves consistency: it keeps the velocity queries on the proper audio manifold while preserving the semantic residual responsible for the edit.
The remaining supplementary sections formalize this claim through definitions, coupling comparisons, contraction results, and empirical evidence.

\section{From Dynamic OT to the Audio Editing Field}
\label{app:dynamic_ot_audio}

\paragraph{Standing primitives.}
Let $\mathcal{X}=\mathbb{R}^{d}$ denote the latent audio space induced by the audio autoencoder, and let $c_{\src}$ and $c_{\tar}$ be the source and target text conditions.
We write $p_0(\cdot\mid c)$ for the clean-latent distribution associated with condition $c$, where the subscript $0$ follows the rectified-flow convention that $t=0$ is the clean endpoint and $t=1$ is the Gaussian endpoint.
For a given real source latent $\vect{x}\in\mathcal{X}$, text-guided editing does not seek an arbitrary draw from $p_0(\cdot\mid c_{\tar})$.
Instead, it seeks a source-conditioned target law, denoted
\begin{equation}
    \label{eq:app_target_law}
    \mu_{\tar|\src}(\cdot\mid \vect{x},c_{\src},c_{\tar}),
\end{equation}
whose mass lies on target-prompt-compatible audio while retaining source attributes not contradicted by the prompt change.
This law is unknown; the role of the pretrained flow is to provide observable local measurements from which an editing field can be estimated.

\subsection{Definitions Used in the Derivation}

\paragraph{(D1) Source-anchored editing density path.}
An ideal edit is represented by a probability path
\begin{equation}
    \label{eq:app_edit_density_path}
    \begin{aligned}
    &\{\rho_\tau\}_{\tau\in[0,1]},
    \qquad
    \rho_0=\delta_{\vect{x}},
    \\
    &\rho_1=\mu_{\tar|\src}(\cdot\mid \vect{x},c_{\src},c_{\tar}).
    \end{aligned}
\end{equation}
where $\tau$ denotes editing progress.
We reserve $t\in[0,1]$ for the rectified-flow noise level.
This separation is important: $\tau$ indexes the unknown semantic edit path, whereas $t$ indexes the noisy states at which the pretrained flow can be queried.

\paragraph{(D2) Editing vector field.}
The complete source-anchored editing field is a time-indexed vector field
\begin{equation}
    \label{eq:app_edit_field}
    u_\tau:\mathcal{X}\rightarrow\mathbb{R}^{d}
\end{equation}
that transports the density path in \eqref{eq:app_edit_density_path} through the continuity equation
\begin{equation}
    \label{eq:app_continuity}
    \partial_\tau\rho_\tau(\vect{y})
    +
    \nabla_{\vect{y}}\!\cdot\!
    \big(\rho_\tau(\vect{y})u_\tau(\vect{y})\big)
    =
    0,
    \qquad
    \tau\in(0,1).
\end{equation}
The field $u_\tau$ is the object that would be used by an ideal source-preserving editor if the source-conditioned target law were available explicitly.

\paragraph{(D3) Rectified-flow probing kernel.}
The pretrained audio flow provides conditional velocities at noisy states, not direct samples from $\mu_{\tar|\src}$.
For a clean latent anchor $\vect{y}$ and noise level $t$, define the rectified-flow probing kernel
\begin{equation}
    \label{eq:app_rf_kernel}
    K_t(\cdot\mid \vect{y})
    =
    \mathcal{L}\!\left((1-t)\vect{y}+t\vect{\epsilon}\right),
    \qquad
    \vect{\epsilon}\sim\mathcal{N}(0,I),
\end{equation}
where $\mathcal{L}$ denotes the law of a random variable.
The conditional guided velocities exposed by the model are denoted
\begin{equation}
    \label{eq:app_guided_velocities}
    V^{\src}(\vect{z},t)
    =
    V_\theta(\vect{z},t,c_{\src}),
    \qquad
    V^{\tar}(\vect{z},t)
    =
    V_\theta(\vect{z},t,c_{\tar}),
\end{equation}
with classifier-free guidance absorbed into $V_\theta$ for notational clarity.

\paragraph{(D4) Coupled observable residual.}
Given a current edited latent $\vect{y}$, a source noisy state should be drawn from $K_t(\cdot\mid\vect{x})$.
A target-side noisy state must be paired with it through a coupling.
For any joint law
\begin{equation}
    \label{eq:app_general_coupling}
    \pi_t
    \in
    \Pi\!\left(
    K_t(\cdot\mid\vect{x}),
    \nu_t
    \right),
\end{equation}
where $\nu_t$ is a target-side noisy marginal and $\Pi$ denotes the set of couplings with the specified marginals, define the observable residual
\begin{equation}
    \label{eq:app_observable_residual}
    \mathcal{R}_t(\vect{y};\pi_t)
    =
    \mathbb{E}_{(\vect{z}^{\src},\vect{z}^{\tar})\sim\pi_t}
    \left[
    V^{\tar}(\vect{z}^{\tar},t)
    -
    V^{\src}(\vect{z}^{\src},t)
    \right].
\end{equation}
This residual is an observable quantity: it depends only on model velocity evaluations and on the chosen source-target noisy coupling.
The central question is how \eqref{eq:app_observable_residual} should be related to the ideal field $u_\tau$ in \eqref{eq:app_continuity}.

\subsection{Dynamic OT Objective for the Ideal Edit}

Among all density paths satisfying the source and target boundary conditions in \eqref{eq:app_edit_density_path}, the dynamic OT principle selects the minimum-energy transport:
\begin{equation}
    \label{eq:app_dynamic_ot}
    \begin{aligned}
    \min_{\rho,u}\quad
    &
    \int_0^1
    \int_{\mathcal{X}}
    \frac{1}{2}\|u_\tau(\vect{y})\|_2^2
    \,d\rho_\tau(\vect{y})\,d\tau
    \\
    \mathrm{s.t.}\quad
    &
    \partial_\tau\rho_\tau
    +
    \nabla\!\cdot(\rho_\tau u_\tau)=0,
    \\
    &
    \rho_0=\delta_{\vect{x}},
    \qquad
    \rho_1=\mu_{\tar|\src}(\cdot\mid \vect{x},c_{\src},c_{\tar}).
    \end{aligned}
\end{equation}
Equation \eqref{eq:app_dynamic_ot} is not used as a computational prescription.
It is a variational definition of the ideal editing geometry: a valid editor should move toward the target-conditioned manifold while avoiding unnecessary displacement from the source-conditioned structure.
The difficulty is that $\mu_{\tar|\src}$ is unknown and the model does not expose $u_\tau$ directly.
The remaining derivation therefore constructs a local estimator of $u_\tau$ from conditional velocity differences.

\subsection{Local Observability Assumptions}

\paragraph{(A1) Conditional velocity difference as a local measurement.}
At a current latent $\vect{y}$ on the edit path, assume that the observable residual provides a noisy local measurement of the ideal editing field:
\begin{equation}
    \label{eq:app_measurement_model}
    \mathcal{R}_t(\vect{y};\pi_t)
    =
    a_t\,u_\tau(\vect{y})
    +
    b_t(\vect{y};\pi_t)
    +
    \xi_t,
    \qquad
    \mathbb{E}[\xi_t]=0.
\end{equation}
Here $a_t>0$ is a scale factor induced by the flow parameterization, $b_t$ is a coupling-dependent bias, and $\xi_t$ is stochastic measurement noise.
The purpose of the later OT coupling is to control $b_t$ by ensuring that the target-side noisy marginal remains compatible with the rectified-flow geometry around $\vect{y}$.

\paragraph{(A2) Local homogeneity of the edit field.}
Within a small neighborhood of $\vect{y}$ and over the flow-noise window used for measurement, the ideal editing field varies slowly enough that a single vector $u$ can approximate $u_\tau(\vect{y})$.
This is the local analogue of replacing the global problem \eqref{eq:app_dynamic_ot} by a myopic estimator at the current point.

\paragraph{(A3) Minimum local kinetic energy.}
Among fields that agree with the observable measurements, the preferred local field is the one with the smallest kinetic energy.
This assumption transfers the Benamou--Brenier energy principle from the global density path to the local estimator.

\subsection{A Local Variational Estimator}

Let $\Omega\subset[0,1]$ be a set of flow-noise levels used to measure the edit direction, and let $\omega(t)\ge 0$ be an integrable weight function.
For a fixed family of couplings $\pi=\{\pi_t\}_{t\in\Omega}$, define the local objective
\begin{equation}
    \label{eq:app_local_objective}
    \begin{aligned}
    \mathcal{J}_{\vect{y}}(u;\pi)
    &=
    \frac{\lambda}{2}\|u\|_2^2
    \\
    &\quad+
    \frac{1}{2}
    \int_{\Omega}
    \omega(t)
    \big\|
    a_t u-\mathcal{R}_t(\vect{y};\pi_t)
    \big\|_2^2
    dt,
    \quad \lambda\ge 0.
    \end{aligned}
\end{equation}
The first term is the local kinetic-energy regularizer inherited from \eqref{eq:app_dynamic_ot}.
The second term enforces agreement with the observable residuals.
The variable $u$ is not an implementation update; it is the local field estimate that approximates the ideal $u_\tau(\vect{y})$.

\paragraph{Proposition 1.}
Assume $\lambda+\int_\Omega\omega(t)a_t^2dt>0$.
For a fixed coupling family $\pi$, the unique minimizer of \eqref{eq:app_local_objective} is
\begin{equation}
    \label{eq:app_local_minimizer}
    u^\star(\vect{y};\pi)
    =
    \frac{
    \int_{\Omega}\omega(t)a_t\,\mathcal{R}_t(\vect{y};\pi_t)\,dt
    }{
    \lambda+\int_{\Omega}\omega(t)a_t^2\,dt
    }.
\end{equation}

\paragraph{Proof.}
The functional \eqref{eq:app_local_objective} is a strictly convex quadratic in $u$ under the stated condition.
Taking its gradient and setting it equal to zero gives
\begin{equation}
    \label{eq:app_normal_equation}
    \lambda u
    +
    \int_\Omega
    \omega(t)a_t
    \left(
    a_t u-\mathcal{R}_t(\vect{y};\pi_t)
    \right)
    dt
    =
    0.
\end{equation}
Collecting the terms multiplying $u$ yields
\begin{equation}
    \left(
    \lambda+\int_\Omega\omega(t)a_t^2dt
    \right)u
    =
    \int_\Omega
    \omega(t)a_t\,\mathcal{R}_t(\vect{y};\pi_t)dt,
\end{equation}
which gives \eqref{eq:app_local_minimizer}.
\hfill$\square$

\paragraph{Interpretation.}
Equation \eqref{eq:app_local_minimizer} formalizes the editing field as an energy-regularized average of observable conditional-velocity differences.
The remaining ambiguity is entirely geometric: the estimator depends on the coupling family $\pi$ used to form the paired noisy states in \eqref{eq:app_observable_residual}.
If the target-side marginal $\nu_t$ in \eqref{eq:app_general_coupling} deviates from the rectified-flow marginal $K_t(\cdot\mid\vect{y})$, the measurement bias term $b_t(\vect{y};\pi_t)$ in \eqref{eq:app_measurement_model} may become non-negligible.
Thus, dynamic OT motivates a minimum-energy editing field.

\section{Unified Coupling View of Inversion, \method, and \methodplus}
\label{app:coupling_view}

\subsection{Admissible Noisy Couplings}

For a source latent $\vect{x}$ and current edited latent $\vect{y}$, define the source and edited noisy marginals
\begin{equation}
    \label{eq:app_noisy_marginals}
    \mu_t^{\src}=K_t(\cdot\mid\vect{x}),
    \qquad
    \mu_t^{\vect{y}}=K_t(\cdot\mid\vect{y}).
\end{equation}
The local estimator in \eqref{eq:app_local_minimizer} is geometrically faithful only if the paired source and target states are drawn from a coupling
\begin{equation}
    \label{eq:app_admissible_coupling}
    \pi_t\in\Pi(\mu_t^{\src},\mu_t^{\vect{y}}).
\end{equation}
We call \eqref{eq:app_admissible_coupling} the \emph{marginal admissibility condition}.
It states that the source branch must lie on the noisy manifold induced by the source latent, and the target branch must lie on the noisy manifold induced by the current edited latent.
This condition is independent of how the final waveform is obtained; it is a mathematical requirement on the noisy states used to estimate the editing field.

The observable residual in \eqref{eq:app_observable_residual} can therefore be decomposed conceptually into two roles.
The velocity difference supplies semantic information by contrasting $c_{\src}$ and $c_{\tar}$.
The coupling supplies geometric information by specifying where those velocities are evaluated.
When the target-side marginal is not $\mu_t^{\vect{y}}$, the residual is measured off the rectified-flow manifold associated with $\vect{y}$, and the bias term in \eqref{eq:app_measurement_model} need not vanish.

\subsection{Inversion and Noising as Bridge Topologies}

Inversion-based editors do not estimate the editing field through a family of admissible local couplings \eqref{eq:app_admissible_coupling}.
Instead, they build a bridge at a selected noise level.
Let $\eta_{t_m}^{\src}$ denote the distribution reached by transporting the source through a source-conditioned forward process to depth $t_m$.
An inversion editor imposes a diagonal bridge coupling
\begin{equation}
    \label{eq:app_inversion_bridge}
    \gamma_{t_m}^{\mathrm{inv}}
    \in
    \Pi(\eta_{t_m}^{\src},\eta_{t_m}^{\src}),
    \qquad
    \gamma_{t_m}^{\mathrm{inv}}
    \big(\{(\vect{z},\vect{z}):\vect{z}\in\mathcal{X}\}\big)=1,
\end{equation}
and then follows the target-conditioned flow from the shared bridge state.
Noising-based editing is recovered by replacing $\eta_{t_m}^{\src}$ with the rectified-flow corruption law $K_{t_m}(\cdot\mid\vect{x})$.

This bridge construction explains the standard edit-strength trade-off.
At small $t_m$, the bridge remains close to the source, so preservation is strong but the target condition has limited freedom.
At large $t_m$, the target flow has more freedom, but the bridge weakens structural attributes carried by the source.
In both cases, source preservation is mediated by a single noisy state rather than by a direct source-to-target editing field along the current edited latent.

\subsection{\method as a Direct but Unconstrained Coupling}

\method removes the noisy bridge and estimates the field directly from paired source- and target-conditioned velocity observations.
In coupling language, its defining principle is to preserve a shared stochastic probe while translating the target-side observation by the current clean-domain displacement.
At the level of probability laws, this gives a target-side marginal of the form
\begin{equation}
    \label{eq:app_audioedit_target_law}
    \begin{aligned}
    \widetilde{\mu}^{\mathrm{A}}_t(\cdot\mid\vect{x},\vect{y})
    &=
    \mathcal{L}\!\left(
    (1-t)\vect{x}
    +
    t\vect{\epsilon}
    +
    (\vect{y}-\vect{x})
    \right),
    \\
    &\hspace{1.2em}
    \vect{\epsilon}\sim\mathcal{N}(0,I).
    \end{aligned}
\end{equation}
Equation \eqref{eq:app_audioedit_target_law} is a distributional description of the coupling, not an implementation rule.
It makes explicit why the method is direct: the source and target observations are compared under the same random probe, and the clean displacement anchors the target branch to the evolving edit.

However, $\widetilde{\mu}^{\mathrm{A}}_t$ is generally not the admissible target marginal $\mu_t^{\vect{y}}$.
Since the rectified-flow marginal around $\vect{y}$ is
\begin{equation}
    \label{eq:app_target_rf_marginal}
    \mu_t^{\vect{y}}
    =
    \mathcal{L}\!\left((1-t)\vect{y}+t\vect{\epsilon}\right),
\end{equation}
the mean of \eqref{eq:app_audioedit_target_law} differs from the mean of \eqref{eq:app_target_rf_marginal} by
\begin{equation}
    \label{eq:app_audioedit_mean_bias}
    \mathbb{E}_{\widetilde{\mu}^{\mathrm{A}}_t}[\vect{z}]
    -
    \mathbb{E}_{\mu_t^{\vect{y}}}[\vect{z}]
    =
    t(\vect{y}-\vect{x}).
\end{equation}
Moreover, both laws have covariance $t^2I$.
Thus their quadratic Wasserstein discrepancy is exactly
\begin{equation}
    \label{eq:app_audioedit_w2_bias}
    W_2^2\!\left(
    \widetilde{\mu}^{\mathrm{A}}_t,
    \mu_t^{\vect{y}}
    \right)
    =
    t^2\|\vect{y}-\vect{x}\|_2^2.
\end{equation}
This identity formalizes \emph{uncertainty drift}.
The direct estimator is semantically meaningful because it contrasts conditional velocities without inversion, but its target-side measurements are increasingly displaced from the correct noisy marginal as either the noise level $t$ or the accumulated edit magnitude $\|\vect{y}-\vect{x}\|_2$ grows.

\subsection{\methodplus as the OT-Admissible Direct Coupling}

\methodplus retains the direct velocity-difference principle while enforcing marginal admissibility.
At each noise level, it selects the quadratic-cost optimal coupling between the two rectified-flow marginals:
\begin{equation}
    \label{eq:app_audioeditplus_ot}
    \pi_t^{+}
    \in
    \arg\min_{\pi\in\Pi(\mu_t^{\src},\mu_t^{\vect{y}})}
    \mathbb{E}_{\pi}
    \big[
    \|\vect{z}^{\tar}-\vect{z}^{\src}\|_2^2
    \big].
\end{equation}
Because $\mu_t^{\src}$ and $\mu_t^{\vect{y}}$ are Gaussian translations with the same covariance $t^2I$, their Monge optimal coupling pairs equal centered perturbations.
Equivalently, if one starts from the translated target-side law in \eqref{eq:app_audioedit_target_law}, the OT-admissible law is obtained by removing the excess high-noise displacement at the level of distributions:
\begin{equation}
    \label{eq:app_distributional_projection}
    \mathcal{L}\!\left(
    \widetilde{\vect{z}}^{\mathrm{A}}_t
    -
    t(\vect{y}-\vect{x})
    \right)
    =
    \mu_t^{\vect{y}},
    \qquad
    \widetilde{\vect{z}}^{\mathrm{A}}_t
    \sim
    \widetilde{\mu}^{\mathrm{A}}_t.
\end{equation}
The identity \eqref{eq:app_distributional_projection} should be read as a marginal correction statement.
It is not a literal algorithmic update; it states that the displacement responsible for the mismatch in \eqref{eq:app_audioedit_mean_bias} is precisely the component removed by the OT-admissible coupling.

Under $\pi_t^{+}$, the paired displacement has cost
\begin{equation}
    \label{eq:app_plus_pair_cost}
    \mathbb{E}_{\pi_t^{+}}
    \big[
    \|\vect{z}^{\tar}-\vect{z}^{\src}\|_2^2
    \big]
    =
    (1-t)^2\|\vect{y}-\vect{x}\|_2^2.
\end{equation}
By contrast, the direct translated coupling carries the full clean-domain displacement through every noise level, yielding paired displacement $\|\vect{y}-\vect{x}\|_2^2$ while failing the admissibility condition.
The OT coupling therefore supplies two corrections simultaneously: it restores the correct target noisy marginal, and it contracts unnecessary displacement as $t$ approaches the high-noise endpoint.

\subsection{Summary of the Coupling Hierarchy}

The three editing paradigms can now be placed in a common hierarchy.
Inversion and noising methods construct an indirect bridge at a selected noisy depth.
\method replaces the bridge with a direct stochastic velocity-difference field, but its translated target-side marginal is not generally admissible.
\methodplus keeps the direct field and imposes the OT-admissible coupling, thereby aligning semantic transport with rectified-flow geometry.

\section{Energy and Transport Contraction Properties}
\label{app:contraction_properties}

This section proves two complementary properties of the OT-admissible coupling.
First, it contracts the transport energy carried through noisy states.
Second, under mild regularity of the pretrained velocity field, it reduces the bias of the observable residual used to estimate the local editing field.
All statements are formulated at the level of distributions and vector fields; no implementation-specific update rule is used.

\subsection{Transport Energy of a Coupling}

Let
\begin{equation}
    \label{eq:app_delta_def}
    \Delta=\vect{y}-\vect{x}
\end{equation}
denote the current clean-domain edit displacement.
For a coupling family $\pi=\{\pi_t\}_{t\in\Omega}$, define its weighted paired transport energy
\begin{equation}
    \label{eq:app_weighted_transport_energy}
    \mathcal{E}_{\omega}(\pi)
    =
    \int_{\Omega}
    \omega(t)
    \mathbb{E}_{\pi_t}
    \big[
    \|\vect{z}^{\tar}-\vect{z}^{\src}\|_2^2
    \big]dt,
    \qquad
    \omega(t)\ge 0.
\end{equation}
This energy measures how much paired motion is carried by the noisy states used to observe the editing field.
It is not the final edit magnitude; rather, it is the instantaneous geometric cost of the coupling used for velocity measurement.

\subsection{Marginal Admissibility and OT Optimality}

\paragraph{Proposition 2.}
For the rectified-flow marginals
\[
    \mu_t^{\src}
    =
    \mathcal{L}((1-t)\vect{x}+t\vect{\epsilon}),
    \qquad
    \mu_t^{\vect{y}}
    =
    \mathcal{L}((1-t)\vect{y}+t\vect{\epsilon}),
\]
with $\vect{\epsilon}\sim\mathcal{N}(0,I)$, the quadratic-cost optimal coupling in \eqref{eq:app_audioeditplus_ot} is the Monge coupling that pairs identical centered perturbations.
Its paired displacement has deterministic norm
\begin{equation}
    \label{eq:app_ot_displacement}
    \|\vect{z}^{\tar,+}-\vect{z}^{\src}\|_2
    =
    (1-t)\|\Delta\|_2.
\end{equation}

\paragraph{Proof.}
Both $\mu_t^{\src}$ and $\mu_t^{\vect{y}}$ are Gaussian measures with covariance $t^2I$ and means $(1-t)\vect{x}$ and $(1-t)\vect{y}$.
For Gaussian measures with equal covariance, the quadratic-cost optimal transport map is translation by the difference of means.
Therefore the optimal coupling pairs the same centered Gaussian perturbation in both marginals.
The displacement under this coupling is the difference of the means,
\[
    (1-t)\vect{y}-(1-t)\vect{x}
    =
    (1-t)\Delta,
\]
which proves \eqref{eq:app_ot_displacement}.
\hfill$\square$

\subsection{Paired-Cost Contraction}

\paragraph{Proposition 3.}
Let $\pi^{\mathrm{A}}$ denote the direct translated coupling whose target-side marginal is $\widetilde{\mu}^{\mathrm{A}}_t$ from \eqref{eq:app_audioedit_target_law}, and let $\pi^+$ denote the OT-admissible coupling from \eqref{eq:app_audioeditplus_ot}.
Then, for every $t\in[0,1]$,
\begin{equation}
    \label{eq:app_pointwise_energy_compare}
    \begin{aligned}
    \mathbb{E}_{\pi_t^+}
    \|\vect{z}^{\tar}-\vect{z}^{\src}\|_2^2
    &=
    (1-t)^2\|\Delta\|_2^2
    \\
    &\le
    \|\Delta\|_2^2
    =
    \mathbb{E}_{\pi_t^{\mathrm{A}}}
    \|\vect{z}^{\tar}-\vect{z}^{\src}\|_2^2.
    \end{aligned}
\end{equation}
Consequently,
\begin{equation}
    \label{eq:app_integrated_energy_compare}
    \begin{aligned}
    \mathcal{E}_{\omega}(\pi^+)
    &=
    \left(\int_{\Omega}\omega(t)(1-t)^2dt\right)
    \|\Delta\|_2^2
    \\
    &\le
    \left(\int_{\Omega}\omega(t)dt\right)
    \|\Delta\|_2^2
    =
    \mathcal{E}_{\omega}(\pi^{\mathrm{A}}).
    \end{aligned}
\end{equation}
If $\|\Delta\|_2>0$ and $\omega$ has positive mass on any subset of $\Omega$ with $t>0$, the inequality in \eqref{eq:app_integrated_energy_compare} is strict.

\paragraph{Proof.}
The OT cost follows from Proposition~2.
The translated direct coupling carries the clean-domain displacement $\Delta$ between the paired source and target observations at every noise level, so its paired cost is $\|\Delta\|_2^2$.
The pointwise comparison follows from $(1-t)^2\le 1$.
Integrating the pointwise identity with nonnegative weight $\omega$ gives \eqref{eq:app_integrated_energy_compare}.
Strictness holds whenever $(1-t)^2<1$ on a set of positive $\omega$-measure and $\Delta\ne 0$.
\hfill$\square$

\paragraph{Endpoint interpretation.}
At $t=0$, the OT coupling retains the full clean-domain displacement, so the target endpoint remains anchored at $\vect{y}$.
At $t=1$, the paired displacement under the OT coupling vanishes, reflecting that both branches share the same Gaussian endpoint.
Thus the coupling preserves the clean edit while removing unnecessary displacement in high-noise regions where the rectified-flow marginals should coincide.

\subsection{Wasserstein Correction of the Drifted Target Marginal}

\paragraph{Proposition 4.}
The marginal mismatch of the direct translated target law satisfies
\begin{equation}
    \label{eq:app_w2_mismatch_restate}
    W_2\!\left(
    \widetilde{\mu}^{\mathrm{A}}_t,
    \mu_t^{\vect{y}}
    \right)
    =
    t\|\Delta\|_2.
\end{equation}
Moreover, the OT-admissible target marginal has zero mismatch:
\begin{equation}
    \label{eq:app_plus_zero_mismatch}
    W_2\!\left(
    \mu_t^{\vect{y}},
    \mu_t^{\vect{y}}
    \right)
    =
    0.
\end{equation}

\paragraph{Proof.}
The two laws $\widetilde{\mu}^{\mathrm{A}}_t$ and $\mu_t^{\vect{y}}$ are Gaussian translations with equal covariance $t^2I$.
Their means differ by $t\Delta$ from \eqref{eq:app_audioedit_mean_bias}.
For equal-covariance Gaussian measures, the squared $W_2$ distance is the squared Euclidean distance between the means, which gives \eqref{eq:app_w2_mismatch_restate}.
The second identity is immediate because the OT-admissible construction uses $\mu_t^{\vect{y}}$ as its target marginal.
\hfill$\square$

\subsection{Residual Bias Bound}

The previous propositions are purely geometric.
We now connect the geometry to the observable velocity residual.
Assume the target-conditioned velocity is Lipschitz in the noisy latent:
\begin{equation}
    \label{eq:app_lipschitz_assumption}
    \begin{aligned}
    \|V^{\tar}(\vect{z}_1,t)-V^{\tar}(\vect{z}_2,t)\|_2
    &\le
    L_t\|\vect{z}_1-\vect{z}_2\|_2,
    \\
    &\hspace{1.0em}L_t<\infty.
    \end{aligned}
\end{equation}
Let $\mathcal{R}_t^{\mathrm{A}}(\vect{y})$ be the observable residual formed with the direct translated target marginal $\widetilde{\mu}^{\mathrm{A}}_t$, and let $\mathcal{R}_t^{+}(\vect{y})$ be the residual formed with the OT-admissible target marginal $\mu_t^{\vect{y}}$.
Both residuals use the same source marginal $\mu_t^{\src}$.

\paragraph{Proposition 5.}
Under \eqref{eq:app_lipschitz_assumption},
\begin{equation}
    \label{eq:app_residual_bias_bound}
    \left\|
    \mathcal{R}_t^{\mathrm{A}}(\vect{y})
    -
    \mathcal{R}_t^{+}(\vect{y})
    \right\|_2
    \le
    L_t\,t\,\|\Delta\|_2.
\end{equation}

\paragraph{Proof.}
The source-velocity expectation is identical in both residuals because both use the same source marginal $\mu_t^{\src}$.
Therefore the residual difference is the difference between target-velocity expectations under $\widetilde{\mu}^{\mathrm{A}}_t$ and $\mu_t^{\vect{y}}$.
By the Kantorovich--Rubinstein Lipschitz bound and $W_1\le W_2$,
\begin{equation}
    \left\|
    \mathbb{E}_{\widetilde{\mu}^{\mathrm{A}}_t}
    V^{\tar}(\vect{z},t)
    -
    \mathbb{E}_{\mu_t^{\vect{y}}}
    V^{\tar}(\vect{z},t)
    \right\|_2
    \le
    L_t
    W_2\!\left(
    \widetilde{\mu}^{\mathrm{A}}_t,
    \mu_t^{\vect{y}}
    \right).
\end{equation}
Substituting Proposition~4 gives \eqref{eq:app_residual_bias_bound}.
\hfill$\square$

\subsection{Local Field Bias Bound}

Finally, we propagate the residual bound through the local variational estimator in \eqref{eq:app_local_minimizer}.
Let
\begin{equation}
    \label{eq:app_estimator_denominator}
    D_{\Omega}
    =
    \lambda+\int_{\Omega}\omega(t)a_t^2dt,
    \qquad
    D_{\Omega}>0.
\end{equation}
Let $u_{\mathrm{A}}^\star(\vect{y})$ and $u_+^\star(\vect{y})$ denote the minimizers obtained from \eqref{eq:app_local_minimizer} using $\mathcal{R}^{\mathrm{A}}$ and $\mathcal{R}^{+}$, respectively.

\paragraph{Proposition 6.}
If \eqref{eq:app_lipschitz_assumption} holds on $\Omega$, then
\begin{equation}
    \label{eq:app_field_bias_bound}
    \|u_{\mathrm{A}}^\star(\vect{y})-u_+^\star(\vect{y})\|_2
    \le
    \frac{
    \int_{\Omega}
    \omega(t)|a_t|L_t\,t\,dt
    }{
    D_{\Omega}
    }
    \|\Delta\|_2.
\end{equation}

\paragraph{Proof.}
Subtracting the two instances of \eqref{eq:app_local_minimizer} gives
\begin{equation}
    u_{\mathrm{A}}^\star(\vect{y})-u_+^\star(\vect{y})
    =
    \frac{
    \int_{\Omega}
    \omega(t)a_t
    \big(
    \mathcal{R}^{\mathrm{A}}_t(\vect{y})
    -
    \mathcal{R}^{+}_t(\vect{y})
    \big)dt
    }{D_{\Omega}}.
\end{equation}
Taking norms and applying Proposition~5 yields \eqref{eq:app_field_bias_bound}.
\hfill$\square$

\paragraph{Consequence.}
The direct stochastic estimator and the OT-admissible estimator agree when $\Delta=0$, as expected at initialization.
As the edit grows, the possible bias of the direct translated coupling scales with the accumulated clean displacement and with the flow-noise factor $t$.
The OT coupling removes the marginal mismatch term itself, so the bound identifies the precise geometric source of uncertainty drift rather than attributing it to stochastic sampling alone.

\section{Gap to the Ideal Editing Field}
\label{app:gap_to_ideal_field}

The preceding sections identify the coupling geometry that makes the observable residual compatible with the rectified-flow marginal.
We now make explicit how the estimated local field can differ from the ideal dynamic-OT field in \eqref{eq:app_continuity}.
This decomposition separates four effects: local approximation error, model error, coupling bias, and stochastic measurement noise.

\subsection{Error Decomposition}

Fix a point $\vect{y}$ on the editing path and write
\begin{equation}
    \label{eq:app_ideal_local_u}
    u=u_\tau(\vect{y})
\end{equation}
for the ideal source-anchored editing field at that point.
For a coupling family $\pi$, refine the measurement model in \eqref{eq:app_measurement_model} as
\begin{equation}
    \label{eq:app_refined_measurement}
    \mathcal{R}^{\pi}_t(\vect{y})
    =
    a_tu
    +
    e_t^{\mathrm{loc}}
    +
    e_t^{\mathrm{mod}}
    +
    e_t^{\mathrm{cpl}}(\pi)
    +
    \xi_t.
\end{equation}
Here $e_t^{\mathrm{loc}}$ is the local-homogeneity error from replacing the true field around $\vect{y}$ by a single vector, $e_t^{\mathrm{mod}}$ is the error induced by using the pretrained conditional velocity as an observable surrogate, $e_t^{\mathrm{cpl}}(\pi)$ is the coupling-induced bias, and $\xi_t$ is zero-mean stochastic measurement noise.
For the OT-admissible coupling, the marginal-drift component of $e_t^{\mathrm{cpl}}$ vanishes.
For the direct translated coupling, Proposition~5 bounds this component by $L_t t\|\Delta\|_2$.

Let $u_\pi^\star(\vect{y})$ denote the minimizer in \eqref{eq:app_local_minimizer} formed with residuals $\mathcal{R}^{\pi}_t$.
Recall $D_\Omega$ from \eqref{eq:app_estimator_denominator}.
For compactness, define the aggregate measurement error
\begin{equation}
    \label{eq:app_aggregate_error}
    e_t^\pi
    =
    e_t^{\mathrm{loc}}
    +
    e_t^{\mathrm{mod}}
    +
    e_t^{\mathrm{cpl}}(\pi)
    +
    \xi_t.
\end{equation}

\paragraph{Proposition 7.}
Under the refined measurement model \eqref{eq:app_refined_measurement}, the gap between the local estimator and the ideal editing field decomposes as
\begin{equation}
    \label{eq:app_gap_decomposition}
    \begin{aligned}
    u_\pi^\star(\vect{y})-u
    &=
    -\frac{\lambda}{D_\Omega}u
    \\
    &\quad+
    \frac{1}{D_\Omega}
    \int_\Omega
    \omega(t)a_t
    e_t^\pi dt.
    \end{aligned}
\end{equation}

\paragraph{Proof.}
Substitute \eqref{eq:app_refined_measurement} into the closed-form estimator \eqref{eq:app_local_minimizer}.
The contribution of the ideal field is
\[
    \frac{\int_\Omega\omega(t)a_t^2dt}{D_\Omega}u
    =
    \left(1-\frac{\lambda}{D_\Omega}\right)u.
\]
Subtracting $u$ gives the regularization term $-\lambda u/D_\Omega$, and the remaining terms yield the integral in \eqref{eq:app_gap_decomposition}.
\hfill$\square$

\paragraph{Corollary 1.}
Taking expectations over the measurement noise and applying the triangle inequality gives
\begin{equation}
    \label{eq:app_expected_gap_bound}
    \begin{aligned}
    \mathbb{E}\|u_\pi^\star(\vect{y})-u\|_2
    &\le
    \frac{\lambda}{D_\Omega}\|u\|_2
    \\
    &\quad+
    \frac{1}{D_\Omega}
    \int_\Omega
    \omega(t)|a_t|
    \\
    &\qquad\cdot
    \Big(
    \ell_t+m_t+c_t^\pi+s_t
    \Big)dt,
    \end{aligned}
\end{equation}
where
\[
    \begin{aligned}
    \ell_t&=\|e_t^{\mathrm{loc}}\|_2,
    &m_t&=\|e_t^{\mathrm{mod}}\|_2,\\
    c_t^\pi&=\|e_t^{\mathrm{cpl}}(\pi)\|_2,
    &s_t&=\mathbb{E}\|\xi_t\|_2.
    \end{aligned}
\]
For \methodplus, the marginal-drift portion of $c_t^\pi$ is zero.
For the original direct translated coupling, this portion is at most $L_t t\|\Delta\|_2$ under \eqref{eq:app_lipschitz_assumption}.

\subsection{Consistency of the OT-Admissible Local Field}

The bound above yields a short consistency statement.
Consider a sequence of local estimators indexed by $h$, with measurement windows $\Omega_h$, weights $\omega_h$, regularization $\lambda_h$, and denominator $D_h$.
Assume:
\begin{enumerate}
    \item $0<c\le |a_t|\le C<\infty$ on every $\Omega_h$.
    \item $D_h>0$ and $\lambda_h/D_h\rightarrow 0$.
    \item The local and model errors vanish in weighted average:
    \[
    \frac{1}{D_h}
    \int_{\Omega_h}
    \omega_h(t)|a_t|
    \big(\ell_t+m_t\big)dt
    \rightarrow 0.
    \]
    \item The stochastic measurement error vanishes in weighted average:
    \[
    \frac{1}{D_h}
    \int_{\Omega_h}
    \omega_h(t)|a_t|s_tdt
    \rightarrow 0.
    \]
\end{enumerate}

\paragraph{Proposition 8.}
Under the assumptions above, the OT-admissible local estimator satisfies
\begin{equation}
    \label{eq:app_consistency_result}
    \mathbb{E}\|u_+^\star(\vect{y})-u_\tau(\vect{y})\|_2
    \rightarrow 0.
\end{equation}

\paragraph{Proof.}
Apply Corollary~1 with $\pi=\pi^+$.
For the OT-admissible coupling, the marginal-drift component of the coupling error is zero by construction.
The remaining terms in \eqref{eq:app_expected_gap_bound} vanish by the four assumptions, which proves \eqref{eq:app_consistency_result}.
\hfill$\square$

\paragraph{Implication for uncertainty drift.}
The consistency statement does not assert that a pretrained audio model is an exact oracle for the unknown source-conditioned target law.
Rather, it states that, once local approximation, model-surrogate error, and stochastic measurement noise are controlled, the OT-admissible coupling does not introduce an additional marginal-drift error.
The original direct translated coupling has an extra term proportional to $t\|\Delta\|_2$ in the residual-bias bound, so its consistency would additionally require this term to vanish.
That requirement is generally incompatible with substantial edits at nonzero flow noise levels, which explains why \method can drift while \methodplus remains geometrically aligned.

\section{Full Algorithm}
\label{app:full_algorithm}

Algorithm~\ref{alg:EchoEditPlusFull} gives the complete practical version of \methodplus corresponding to the simplified main-paper algorithm.
The main text suppresses three practical controls for clarity: Monte Carlo averaging of the local velocity residual, the edit window defined by $(\nmax,\nmin)$, and an optional target-conditioned low-noise refinement.
These controls do not change the coupling principle.
The source and target branches are still evaluated through the OT-admissible noisy coupling; the additional parameters only determine how strongly and over which part of the flow trajectory this field is applied.
We write $V_w(\cdot,t,c)$ for the classifier-free-guided conditional velocity under prompt $c$ and guidance scale $w$.

\begin{algorithm}[t]
    \caption{\textbf{Full \methodplus algorithm}}
    \label{alg:EchoEditPlusFull}
    \textbf{Input}: source latent $X^{\mathrm{src}}$, prompts $c_{\mathrm{src}},c_{\mathrm{tar}}$, schedule $\{t_i\}_{i=0}^{T}$, edit window $\nmax,\nmin$, averaging count $\navg$, guidance scales $w_{\mathrm{src}},w_{\mathrm{tar}}$, step scale $\eta$.\\
    \textbf{Output}: edited latent $X^{\mathrm{tar}}$.
    \begin{algorithmic}[1]
    \State $Z^{+}_{t_{\nmax}}\leftarrow X^{\mathrm{src}}$
    \For{$i=\nmax$ \textbf{to} $\nmin+1$}
        \State $\bar{V}_{t_i}\leftarrow 0$
        \For{$k=1$ \textbf{to} $\navg$}
            \State Sample $\epsilon_{i,k}\sim\mathcal{N}(0,I)$
            \State $\hat{Z}^{\src}_{t_i,k}\leftarrow (1-t_i)X^{\mathrm{src}}+t_i\epsilon_{i,k}$
            \State $\hat{Z}^{\tar,+}_{t_i,k}\leftarrow (1-t_i)Z^{+}_{t_i}+t_i\epsilon_{i,k}$
            \State $V^{\src}_{t_i,k}\leftarrow V_{w_{\mathrm{src}}}(\hat{Z}^{\src}_{t_i,k},t_i,c_{\mathrm{src}})$
            \State $V^{\tar}_{t_i,k}\leftarrow V_{w_{\mathrm{tar}}}(\hat{Z}^{\tar,+}_{t_i,k},t_i,c_{\mathrm{tar}})$
            \State $\bar{V}_{t_i}\leftarrow \bar{V}_{t_i}+\big(V^{\tar}_{t_i,k}-V^{\src}_{t_i,k}\big)/\navg$
        \EndFor
        \State $Z^{+}_{t_{i-1}}\leftarrow Z^{+}_{t_i}+\eta(t_{i-1}-t_i)\bar{V}_{t_i}$
    \EndFor
    \If{$\nmin>0$}
        \State Sample $\epsilon\sim\mathcal{N}(0,I)$
        \State $X_{t_{\nmin}}\leftarrow (1-t_{\nmin})Z^{+}_{t_{\nmin}}+t_{\nmin}\epsilon$
        \For{$i=\nmin$ \textbf{to} $1$}
            \State $V^{\tar}_{t_i}\leftarrow V_{w_{\mathrm{tar}}}(X_{t_i},t_i,c_{\mathrm{tar}})$
            \State $X_{t_{i-1}}\leftarrow X_{t_i}+\eta(t_{i-1}-t_i)V^{\tar}_{t_i}$
        \EndFor
        \State \textbf{return} $X^{\mathrm{tar}}=X_{0}$
    \Else
        \State \textbf{return} $X^{\mathrm{tar}}=Z^{+}_{0}$
    \EndIf
    \end{algorithmic}
\end{algorithm}

\section{Experiment Configuration and Details}
\label{app:exp_config}

\subsection{Default Configuration}

All experiments are conducted on a single 24GB NVIDIA RTX 3090 GPU. All experiments use Stable Audio 3 medium~\cite{evans2026stableaudio3} with the SAME latent autoencoder~\cite{parker2026samesemanticallyalignedmusicautoencoder}.
Waveforms are encoded and decoded at 44.1 kHz; the latent tensor has shape $(B,32,T_{\mathrm{lat}})$.
Text conditioning follows the T5+CLAP conditioning pipeline.
Unless otherwise stated, \methodplus adopts $T=28$ solver steps, $\nmax=24$, $\nmin=0$, $\navg=1$, source guidance scale $w_{\src}=1.5$, target guidance scale $w_{\tar}=4.5$, and step coefficient $\eta=1.0$.
Sound-effect clips use their annotated durations with 4.0 seconds of padding; music examples use 30-second excerpts for the main comparison, with long-duration variants spanning 15--120 seconds. For long-form music clips (${\ge}75$ seconds), padding is increased to 6.0 seconds to preserve temporal context.

\begin{table}[htpb]
    \centering
    \small
    \begin{tabular}{@{}ll@{}}
        \toprule
        Parameter & Value \\
        \midrule
        Backbone & Stable Audio 3 medium \\
        Autoencoder & SAME \\
        Sampling rate & 44.1 kHz \\
        Solver steps $T$ & 28 \\
        Edit window $\nmax$ & 24 \\
        Style control $\nmin$ & 0 \\
        Velocity averaging $\navg$ & 1 \\
        Source guidance $w_{\src}$ & 1.5 \\
        Target guidance $w_{\tar}$ & 4.5 \\
        Step coefficient $\eta$ & 1.0 \\
        Duration padding & 4.0 s \\
        CFG rescaling & enabled \\
        \bottomrule
    \end{tabular}
    \caption{Default \methodplus configuration.}
    \label{tab:default_config}
\end{table}

\subsection{Hyperparameters for All Compared Methods}
\label{app:all_hyperparams}

We compare \methodplus against four baseline families on the same Stable Audio 3 backbone: ODE inversion, SDEdit~\cite{meng2022sdeditguidedimagesynthesis}, FireFlow~\cite{deng2024fireflowfastinversionrectified}, and the original \method with the displacement coupling~\cite{kulikov2025floweditinversionfreetextbasedediting}.
All methods share the same source latents, prompt pairs, duration metadata, and random seeds.
Tables~\ref{tab:hyperparams-echoedit}--\ref{tab:hyperparams-fireflow} list the complete hyperparameters for each method.
Bold entries indicate the default configuration used for the main results in Table~1 of the main text.
For the Pareto analysis in Fig.~3 of the main text, target guidance $w_{\tar}$ is swept for \methodplus, \method, ODE inversion, and FireFlow, while the noise strength is swept for SDEdit.

\begin{table}[htpb]
    \centering
    \small
    \begin{tabular}{@{}ccccccc@{}}
        \toprule
        & $T$ & $\nmax$ & $\nmin$ & $\navg$ & $w_{\src}$ & $w_{\tar}$ \\
        \midrule
        \method & 28 & 24 & 0 & 1 & 1.5 & 4.5 \\
        \methodplus & 28 & 24 & 0 & 1 & 1.5 & 3.5,\,\textbf{4.5},\,5.5 \\
        \bottomrule
    \end{tabular}
    \caption{\textbf{\method / \methodplus hyperparameters.}}
    \label{tab:hyperparams-echoedit}
\end{table}

\begin{table}[htpb]
    \centering
    \small
    \begin{tabular}{@{}ccccc@{}}
        \toprule
        & $T$ & $\nmax$ & $w_{\src}$ & $w_{\tar}$ \\
        \midrule
        ODE Inv. & 28 & \textbf{24} & 1.5 & 2.5,\,3.5,\,\textbf{4.5},\,5.5,\,6.5 \\
        \bottomrule
    \end{tabular}
    \caption{\textbf{ODE inversion hyperparameters.}}
    \label{tab:hyperparams-odeinv}
\end{table}

\begin{table}[htpb]
    \centering
    \small
    \begin{tabular}{@{}ccccc@{}}
        \toprule
        & $T$ & strength $s$ & $w_{\tar}$ \\
        \midrule
        SDEdit & 28 & \textbf{0.50},\,0.60,\,0.70,\,0.80,\,0.90 & 4.5 \\
        \bottomrule
    \end{tabular}
    \caption{\textbf{SDEdit hyperparameters.}}
    \label{tab:hyperparams-sdedit}
\end{table}

\begin{table}[htpb]
    \centering
    \small
    \begin{tabular}{@{}ccccc@{}}
        \toprule
        & $T$ & $\nmax$ & $w_{\src}$ & $w_{\tar}$ \\
        \midrule
        FireFlow & 28 & 24 & 1.5 & 2.5,\,3.5,\,\textbf{4.5},\,5.5,\,6.5 \\
        \bottomrule
    \end{tabular}
    \caption{\textbf{FireFlow hyperparameters.}}
    \label{tab:hyperparams-fireflow}
\end{table}


For SDEdit, the noise strength $s$ determines the fraction of the total schedule used for corruption: $s=0.50$ adds noise up to $t=0.50$, producing a moderate edit that largely retains source structure; $s=0.90$ adds noise up to $t=0.90$, granting the target condition greater generative freedom at the cost of source structure.

\subsection{Baseline Algorithm Pseudocode}
\label{app:baseline_pseudocode}

For completeness, we provide the pseudocode of the baseline editing algorithms evaluated in our experiments.
All algorithms share the ascending noise schedule $0=t_0<t_1<\cdots<t_T=1$, consistent with the rectified-flow convention where $t=0$ is the clean-audio endpoint and $t=1$ is the Gaussian endpoint.
The pretrained conditional velocity field under classifier-free guidance scale $w$ and text condition $c$ is denoted $V_{w}(\vect{z},t,c)$.
The source and target prompts are written $c_{\src}$ and $c_{\tar}$; the source latent is $\vect{x}^{\src}$.

\begin{algorithm}[htpb]
    \caption{\textbf{SDEdit} --- Single-pass noising and denoising.}
    \label{alg:sdedit}
    \begin{algorithmic}[1]
    \State {\bfseries Input:} source latent $\vect{x}^{\src}$, target prompt $c_{\tar}$, schedule $\{t_i\}_{i=0}^{T}$, edit strength $s$, target guidance $w_{\tar}$.
    \State {\bfseries Output:} edited latent $\vect{x}^{\tar}$.
    \State $m \leftarrow \mathrm{round}\big((1-s)\,T\big)$
    \State Sample $\vect{\epsilon}\sim\mathcal{N}(0,\vect{I})$
    \State $\vect{z}_{t_m} \leftarrow (1-t_m)\,\vect{x}^{\src} + t_m\,\vect{\epsilon}$
    \For{$i=m$ {\bfseries down to} $1$}
        \State $\vect{v} \leftarrow V_{w_{\tar}}(\vect{z}_{t_i},\,t_i,\,c_{\tar})$
        \State $\vect{z}_{t_{i-1}} \leftarrow \vect{z}_{t_i} + (t_{i-1}-t_i)\,\vect{v}$
    \EndFor
    \State \Return $\vect{z}_{t_0}$
    \end{algorithmic}
\end{algorithm}

\begin{algorithm}[htpb]
    \caption{\textbf{ODE Inversion} --- Two-pass source-inversion editing.}
    \label{alg:odeinversion}
    \begin{algorithmic}[1]
    \State {\bfseries Input:} source latent $\vect{x}^{\src}$, source prompt $c_{\src}$, target prompt $c_{\tar}$, schedule $\{t_i\}_{i=0}^{T}$, edit depth $\nmax$, guidance scales $w_{\src},w_{\tar}$.
    \State {\bfseries Output:} edited latent $\vect{x}^{\tar}$.
    \State $\vect{z}^{\src}_{t_0} \leftarrow \vect{x}^{\src}$
    \For{$i=1$ {\bfseries to} $\nmax$}
        \State $\vect{v} \leftarrow V_{w_{\src}}(\vect{z}^{\src}_{t_{i-1}},\,t_{i-1},\,c_{\src})$
        \State $\vect{z}^{\src}_{t_i} \leftarrow \vect{z}^{\src}_{t_{i-1}} + (t_i-t_{i-1})\,\vect{v}$
    \EndFor
    \State $\vect{z}^{\tar}_{t_{\nmax}} \leftarrow \vect{z}^{\src}_{t_{\nmax}}$
    \For{$i=\nmax$ {\bfseries down to} $1$}
        \State $\vect{v} \leftarrow V_{w_{\tar}}(\vect{z}^{\tar}_{t_i},\,t_i,\,c_{\tar})$
        \State $\vect{z}^{\tar}_{t_{i-1}} \leftarrow \vect{z}^{\tar}_{t_i} + (t_{i-1}-t_i)\,\vect{v}$
    \EndFor
    \State \Return $\vect{z}^{\tar}_{t_0}$
    \end{algorithmic}
\end{algorithm}

\begin{algorithm}[htpb]
    \caption{\textbf{FireFlow} --- Second-order midpoint inversion with velocity reuse.}
    \label{alg:fireflow}
    \begin{algorithmic}[1]
    \State {\bfseries Input:} source latent $\vect{x}^{\src}$, source prompt $c_{\src}$, target prompt $c_{\tar}$, schedule $\{t_i\}_{i=0}^{T}$, edit depth $\nmax$, guidance scales $w_{\src},w_{\tar}$.
    \State {\bfseries Output:} edited latent $\vect{x}^{\tar}$.
    \Function{MidStep}{$\vect{z},\,t_a,\,t_b,\,c,\,w,\,\vect{v}_{\mathrm{p}}$}
        \State $h \leftarrow t_b - t_a$
        \If{$\vect{v}_{\mathrm{p}} = \varnothing$}
            \State $\vect{v}_{\mathrm{p}} \leftarrow V_{w}(\vect{z},\,t_a,\,c)$
        \EndIf
        \State $\vect{z}_{\mathrm{mid}} \leftarrow \vect{z} + \frac{1}{2}h\,\vect{v}_{\mathrm{p}}$
        \State $\vect{v}_{\mathrm{mid}} \leftarrow V_{w}(\vect{z}_{\mathrm{mid}},\,t_a+\frac{1}{2}h,\,c)$
        \State \Return $\big(\vect{z} + h\,\vect{v}_{\mathrm{mid}},\; \vect{v}_{\mathrm{mid}}\big)$
    \EndFunction
    \State $\vect{z} \leftarrow \vect{x}^{\src}$; \quad $\vect{v}_{\mathrm{p}} \leftarrow \varnothing$
    \For{$i=1$ {\bfseries to} $\nmax$}
        \State $(\vect{z},\,\vect{v}_{\mathrm{p}}) \leftarrow \textsc{MidStep}(\vect{z},\,t_{i-1},\,t_i,\,c_{\src},\,w_{\src},\,\vect{v}_{\mathrm{p}})$
    \EndFor
    \State $\vect{v}_{\mathrm{p}} \leftarrow \varnothing$ \Comment{velocity field changes with the prompt}
    \For{$i=\nmax$ {\bfseries down to} $1$}
        \State $(\vect{z},\,\vect{v}_{\mathrm{p}}) \leftarrow \textsc{MidStep}(\vect{z},\,t_i,\,t_{i-1},\,c_{\tar},\,w_{\tar},\,\vect{v}_{\mathrm{p}})$
    \EndFor
    \State \Return $\vect{z}$
    \end{algorithmic}
\end{algorithm}

Algorithm~\ref{alg:sdedit} describes SDEdit, which directly corrupts the source to an intermediate noise level $t_m$ determined by the edit strength $s$ and denoises under the target condition alone.
Algorithm~\ref{alg:odeinversion} describes ODE inversion, which first integrates the source-conditioned velocity forward to depth $\nmax$, then initializes the reverse trajectory at the resulting noisy state and integrates the target-conditioned velocity backward to the clean endpoint.
Algorithm~\ref{alg:fireflow} follows the same two-pass topology but replaces first-order Euler integration with a midpoint (second-order) corrector.
By caching the midpoint velocity as the initial estimate of the next step, FireFlow achieves $\mathcal{O}(\Delta t^3)$ local truncation error at one network function evaluation per step after the first, matching the NFE budget of Euler while approaching the accuracy of a two-evaluation Heun scheme.
The velocity cache is reset between the source and target passes because the velocity field changes with the conditioning signal.

By contrast, \method and \methodplus (Algorithm~\ref{alg:EchoEditPlusFull}) avoid both the forward source integration and the noisy bridge entirely.
Instead of reconstructing a source trajectory into noise and then regenerating under the target prompt, they construct an editing field directly from paired source- and target-conditioned velocity evaluations at matched noise levels.
The two methods differ only in the coupling used to form the noisy state pairs: \method displaces the target-side probe by the full clean-domain edit residual, while \methodplus uses the OT-admissible coupling $\pi_t^{+}$ that aligns both probes with their respective rectified-flow marginals.
This difference in coupling geometry, rather than any change to the semantic objective, is what suppresses uncertainty drift in \methodplus.

\section{Supplementary Experimental Evidence}
\label{app:supplementary_experiments}

The preceding derivation suggests four empirical signatures.
First, varying the amount of transport should trace a smooth semantic--preservation frontier instead of producing irregular latent perturbations.
Second, once the coupling is geometrically admissible, stochastic averaging should act mainly as variance reduction, not as the primary source of consistency.
Third, the same preservation mechanism should persist across edit operations, prompt perturbations, seeds, and sequential edits.
Fourth, failures should concentrate in cases where the requested target does not define a sharply constrained source-conditioned target law.

\subsection{Synthetic Transport-Cost Diagnostic Against Inversion}
\label{app:transport_diagnostic}

The main text uses the transport-cost diagnostic to separate the geometry of an editing path from reconstruction artifacts that can occur on real recordings.
Here the source audio is itself generated by Stable Audio 3 from text prompts, so the samples lie close to the model's own generative manifold.
In this setting, the comparison is not primarily about whether a real waveform can be reconstructed, but about which source-target pairing is induced once the editor is asked to move a generated source toward a target condition.

The synthetic benchmark contains two prompt families.
For sound effects, we used a large language model to generate 64 synthetic acoustic prompts following the official Stable Audio 3 prompt guide.
These prompts specify concrete sound events together with contextual descriptors such as environment, material quality, temporal density, or acoustic texture, so that the resulting clips have source attributes that should not be erased by the edit.
For music, we directly used 64 natural-language audio descriptions from the Song Describer Dataset~\cite{manco2023songdescriberdatasetcorpus} as synthetic prompts.
Each prompt was rendered into a source clip by Stable Audio 3.

For every generated source clip, we constructed target prompts covering replacement, addition, and deletion.
Replacement substitutes an existing sound event, instrument, style, or production attribute while keeping the rest of the description fixed.
Addition introduces a new acoustic component into the source scene or arrangement.
Deletion removes a specified component while preserving the residual context.
All preservation metrics compare each edited sample with its paired generated source, while target-alignment metrics compare the edit with the target prompt.
The independently generated target set is used only for the FAD column; the uncertainty-drift diagnostic is measured directly along the editing trajectory.

For a direct editor, let $\widetilde{\mu}^{\mathrm{A}}_{t_i}$ denote the target-side noisy marginal actually queried by the translated \method coupling at step $t_i$, and let $\mu^{\vect{y}_i}_{t_i}=K_{t_i}(\cdot\mid\vect{y}_i)$ denote the admissible rectified-flow marginal associated with the current edited latent.
The trajectory-wise marginal mismatch is measured by the RMS-normalized Wasserstein discrepancy
\begin{equation}
    \label{eq:app_empirical_w2_mm}
    d_i^{\mathrm{MM}}
    =
    \frac{1}{\sqrt{D}}
    W_2\!\left(
    \widetilde{\mu}^{\mathrm{A}}_{t_i},
    \mu^{\vect{y}_i}_{t_i}
    \right),
\end{equation}
where $D$ is the latent dimensionality used for normalization across durations.
Since these two marginals have equal covariance under the rectified-flow kernel, \eqref{eq:app_empirical_w2_mm} is the RMS-normalized counterpart of the mean displacement characterized in Proposition~4.
We report the schedule-weighted mean $W_2^{\mathrm{MM}}$ and the maximum value $W_{2,\max}^{\mathrm{MM}}$ over the editing trajectory.
For \methodplus, the OT-admissible target marginal is $\mu^{\vect{y}_i}_{t_i}$ by construction, so both quantities are zero up to numerical precision.

Table~\ref{tab:transport} reports the full diagnostic.
The latent MSE provides the most direct estimate of the transport cost induced by the source-target pairing.
Acoustic distances such as LSD, MCD, and LPAPS measure whether that lower latent cost survives decoding into spectro-temporal structure.
\method exhibits a nonzero marginal mismatch, with mean/max $W_2^{\mathrm{MM}}$ equal to $0.11/0.15$ on sound effects and $0.14/0.19$ on music.
This mismatch is strongly correlated with degradation in the induced source-target pairing: across per-example \method trajectories, the Pearson correlation between $W_2^{\mathrm{MM}}$ and latent MSE is $0.98$ in both domains.
Larger mismatch also tracks weaker decoded preservation, with positive correlation to MCD ($0.53$ on sound effects and $0.31$ on music) and negative correlation to CLAP-A ($-0.32$ and $-0.26$) and structure scores ($-0.39$ and $-0.22$).
\methodplus eliminates the measured marginal mismatch and simultaneously obtains the lowest latent MSE and the best decoded preservation metrics while maintaining target alignment comparable to \method.
This indicates that the OT-admissible coupling reduces uncertainty drift itself rather than merely producing a weaker edit.

\begin{table*}[htpb]
    \centering
    \small
    \setlength{\tabcolsep}{2.9pt}
    \begin{tabular}{llcccccccccc}
        \toprule
        \textbf{Domain} & \textbf{Method} & \textbf{Latent MSE} $\downarrow$ & \textbf{CLAP-T} $\uparrow$ & \textbf{CLAP-A} $\uparrow$ & \textbf{LSD} $\downarrow$ & \textbf{MCD} $\downarrow$ & \textbf{LPAPS} $\downarrow$ & \textbf{Structure} $\uparrow$ & \textbf{FAD} $\downarrow$ & $\boldsymbol{W_2^{\mathrm{MM}}}$ $\downarrow$ & $\boldsymbol{W_{2,\max}^{\mathrm{MM}}}$ $\downarrow$ \\
        \midrule
        \multirow{5}{*}{SFX} & FireFlow & 1.14 & 0.38 & 0.50 & 23.10 & 603.91 & 0.27 & 0.49 & 45.92 & -- & -- \\
        & ODE Inv. & 0.92 & 0.39 & 0.52 & 22.55 & 587.47 & 0.25 & 0.51 & 40.05 & -- & -- \\
        & SDEdit & 1.80 & 0.36 & 0.46 & 24.55 & 664.03 & 0.28 & 0.46 & 55.62 & -- & -- \\
        & \method & 0.41 & \textbf{0.46} & 0.66 & 18.55 & 506.43 & 0.18 & 0.56 & \textbf{31.32} & 0.11 & 0.15 \\
        & \textbf{\methodplus} & \textbf{0.10} & 0.44 & \textbf{0.80} & \textbf{12.75} & \textbf{346.60} & \textbf{0.11} & \textbf{0.63} & 34.83 & \textbf{0.00} & \textbf{0.00} \\
        \midrule
        \multirow{5}{*}{Music} & FireFlow & 1.30 & 0.58 & 0.78 & 16.98 & 529.19 & 0.19 & 0.91 & 30.71 & -- & -- \\
        & ODE Inv. & 1.15 & 0.58 & 0.80 & 16.24 & 512.35 & 0.17 & 0.93 & 27.93 & -- & -- \\
        & SDEdit & 3.09 & 0.48 & 0.64 & 19.01 & 596.27 & 0.29 & 0.87 & 61.97 & -- & -- \\
        & \method & 0.64 & \textbf{0.59} & 0.83 & 14.56 & 429.33 & 0.14 & 0.93 & 24.55 & 0.14 & 0.19 \\
        & \textbf{\methodplus} & \textbf{0.12} & 0.58 & \textbf{0.94} & \textbf{9.69} & \textbf{243.51} & \textbf{0.06} & \textbf{0.99} & \textbf{23.20} & \textbf{0.00} & \textbf{0.00} \\
        \bottomrule
    \end{tabular}
    \caption{Synthetic generated-source diagnostic. Source clips are generated from source prompts, edited under target prompts, and compared with the source or an independently generated target-prompt set. Best values are bold. $W_2^{\mathrm{MM}}$ and $W_{2,\max}^{\mathrm{MM}}$ denote the mean and maximum RMS-normalized $W_2$ marginal mismatch along the direct editing trajectory; ``--'' indicates that inversion/noising methods do not define the corresponding direct target-query marginal.}
    \label{tab:transport}
\end{table*}

\subsection{Benchmark Construction and Edit-Strategy Balance}
\label{app:benchmark_construction}

The main paper reports the benchmark protocol at a high level; here we clarify the dataset construction used for the supplementary analyses.
The evaluation is built around source-conditioned editing rather than unconstrained text-to-audio generation.
Each instance therefore consists of a real source waveform, a source prompt describing the observed audio, and a target prompt that changes a semantically meaningful acoustic attribute while preserving the surrounding temporal or musical context whenever the edit permits it.
This design makes the benchmark compatible with the theoretical object studied in the paper: a source-conditioned target law whose support should remain close to the source manifold except along directions required by the edit.

The sound-effect split is derived from FSD50K~\cite{Fonseca_2022}, a broad collection of real-world acoustic events.
We use this domain to evaluate edits involving environmental scenes, object interactions, percussive transients, and short acoustic textures, where source preservation is largely expressed through event timing, background ambience, and local spectro-temporal structure.
The music split is derived from the Song Describer Dataset~\cite{manco2023songdescriberdatasetcorpus}, whose natural-language captions describe musical excerpts at the level of instrumentation, genre, mood, production character, and ensemble composition.
These captions are converted into prompt pairs that alter a controlled musical attribute while retaining the surrounding musical scene, making the split sensitive to melodic contour, rhythmic continuity, timbral identity, and long-range arrangement structure.
Together, the two domains expose the editor to both localized acoustic transformations and globally organized musical transformations.

This construction is intended to test the central claim of the paper: an editor should move toward the target-conditioned manifold without abandoning the source-conditioned structure.
For this reason, the benchmark emphasizes operation diversity rather than a single dominant edit type.
We group edits into three user-facing strategies: addition, replacement, and deletion.
Addition introduces a new acoustic attribute or musical component into the existing scene; replacement changes the identity, material character, instrument, or style of content already present; deletion suppresses a specified component while keeping the residual scene or arrangement coherent.
These strategies stress different aspects of the transport field.
Addition requires local semantic insertion without diffuse regeneration, replacement tests whether the editor can alter content identity along a low-cost path, and deletion probes whether suppressing target content can be achieved without collapsing the source layout.

\begin{figure}[t]
    \centering
    \begin{subfigure}[t]{0.47\columnwidth}
        \centering
        \includegraphics[width=0.86\linewidth]{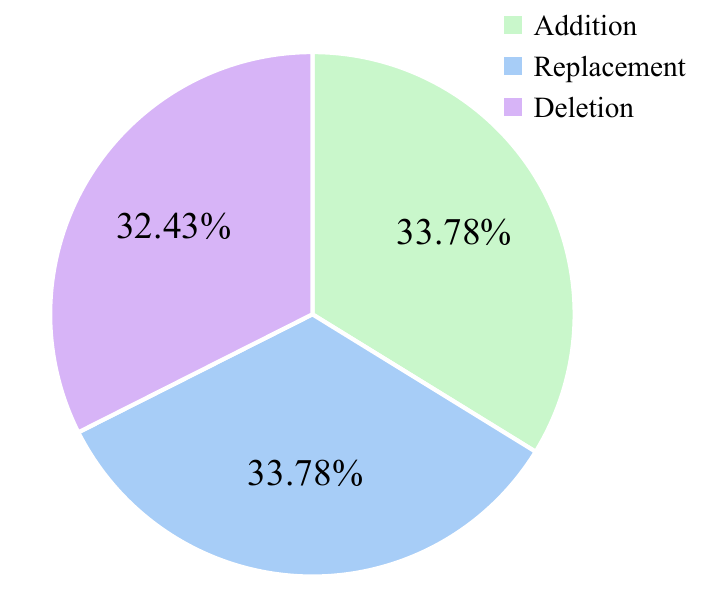}
        \caption{FSD50K sound effects}
        \label{fig:app_sfx_edit_strategy_pie}
    \end{subfigure}
    \hfill
    \begin{subfigure}[t]{0.47\columnwidth}
        \centering
        \includegraphics[width=0.86\linewidth]{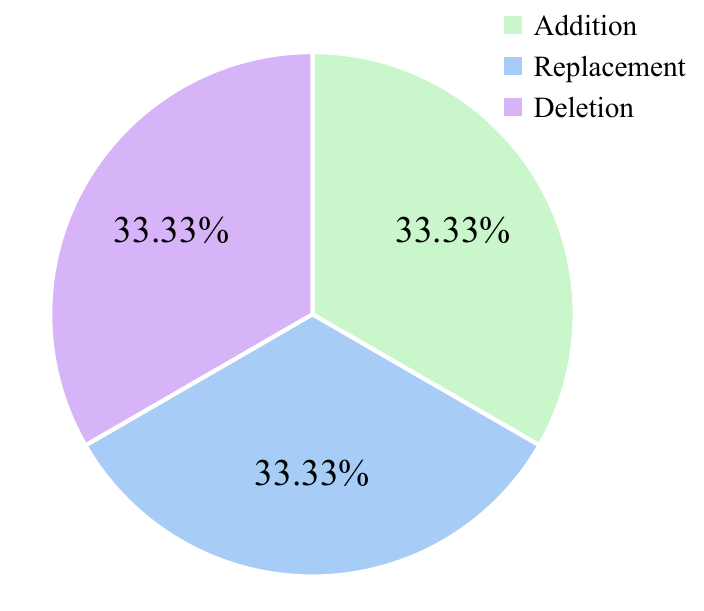}
        \caption{Song Describer music}
        \label{fig:app_music_edit_strategy_pie}
    \end{subfigure}
    \caption{\textbf{Edit-strategy composition of the real-audio benchmark.}
    Both the FSD50K sound-effect split and the Song Describer music split are balanced across addition, replacement, and deletion.
    This construction prevents the evaluation from being dominated by a single operation regime and supports the operation-wise analysis in Table~\ref{tab:operation_app}, where the preservation behavior of \methodplus is examined across distinct source-conditioned target laws.}
    \label{fig:app_edit_strategy_pies}
\end{figure}

Figure~\ref{fig:app_edit_strategy_pies} shows that the two benchmark domains follow a deliberately balanced strategy composition.
The balance is important because uncertainty drift has operation-dependent manifestations.
In sound effects, an over-aggressive replacement can erase ambience or transient placement, an addition can spread energy into unrelated time-frequency regions, and a deletion can degenerate into unrestricted scene regeneration.
In music, the same taxonomy stresses different structural invariants: replacement may disturb instrumentation or genre while preserving rhythm, addition must introduce a new layer without masking the existing arrangement, and deletion must remove a described component without damaging melodic or harmonic continuity.
Consequently, strong aggregate performance is meaningful only if it remains stable under this taxonomy.
The operation-wise results in Sec.~\ref{app:operation_robustness} confirm that the OT-admissible coupling improves preservation across these regimes, supporting the interpretation that \methodplus regularizes the editing geometry rather than exploiting a narrow category bias.

\subsection{Qualitative Energy and Spectro-Temporal Evidence}
\label{app:qualitative_energy}

\begin{figure*}[htpb]
    \centering
    \includegraphics[width=\textwidth]{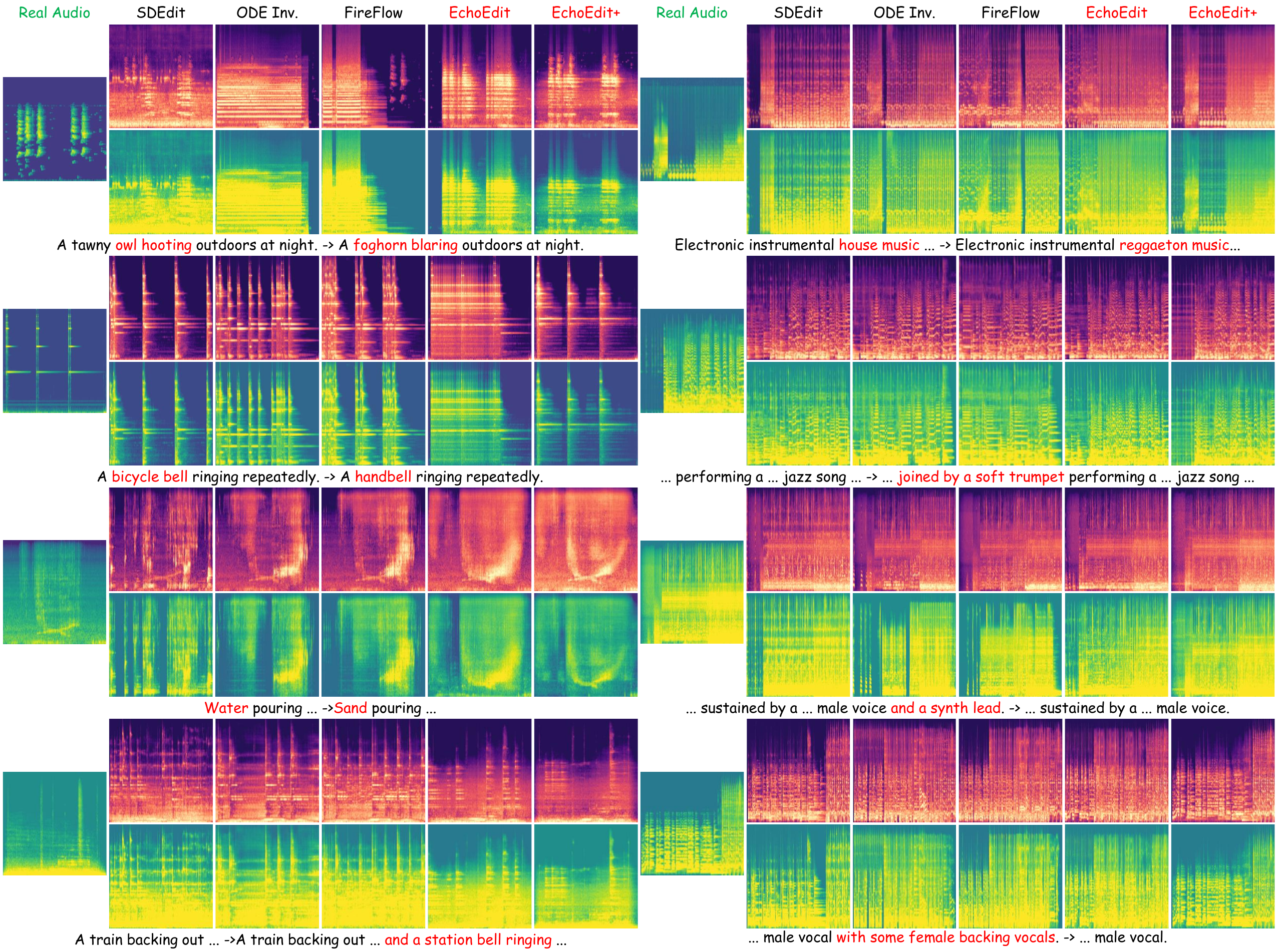}
    \caption{\textbf{Qualitative comparison and editing-effort visualization.}
    We compare inversion/noising baselines, \method, and \methodplus on representative source-to-target audio edits.
    Inversion-style methods often introduce broad spectro-temporal reorganization because the edit is mediated by a noisy bridge.
    \method avoids this bridge but can still spend effort diffusely when the translated stochastic coupling drifts away from the admissible noisy marginal.
    \methodplus yields a more source-consistent edit by regularizing the direct residual through the OT-admissible coupling, preserving non-target temporal structure while modifying the intended auditory content.}
    \label{fig:app_energy}
\end{figure*}

The contraction result in Sec.~\ref{app:contraction_properties} is a statement about the coupling used to measure the editing field: an admissible OT coupling carries less unnecessary paired motion through noisy states than the translated direct coupling.
Figure~\ref{fig:app_energy} provides a qualitative counterpart to this claim.
For each edit, we show the source spectrogram, edited spectrograms from representative baselines, and the corresponding editing-effort visualizations.
The effort maps should be read as diagnostics of where the editor spends transport energy in the spectro-temporal plane, not as separate training objectives.

The visual pattern is consistent with the formal energy analysis.
Inversion and noising methods obtain editability by passing through high-noise states, which can alter background texture and event timing even when the target concept is plausible.
The original \method replaces this bridge with direct conditional transport, but its unconstrained translated coupling can still create off-manifold measurements.
\methodplus changes the geometry of the measurement rather than the semantic objective: by coupling $K_t(\cdot\mid\vect{x}^{\src})$ and $K_t(\cdot\mid\vect{y})$ at minimum quadratic cost, it reduces high-noise displacement and yields edits whose effort is more aligned with the requested acoustic change.

\subsection{Distributional Search-Space Visualization}
\label{app:distributional_visualization}

\begin{figure*}[htpb]
    \centering
    \begin{subfigure}[t]{0.24\textwidth}
        \includegraphics[width=\linewidth]{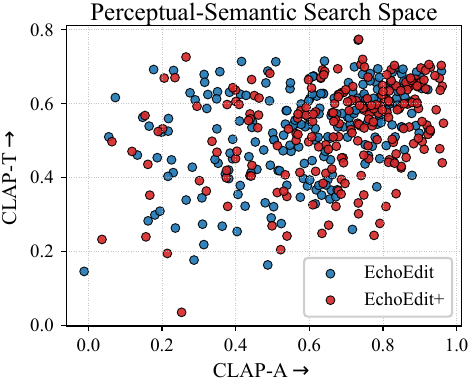}
        \caption{SFX: CLAP-A}
    \end{subfigure}
    \begin{subfigure}[t]{0.24\textwidth}
        \includegraphics[width=\linewidth]{figures/audioedit_comparision/search_space_sfx_lpaps.pdf}
        \caption{SFX: LPAPS}
    \end{subfigure}
    \begin{subfigure}[t]{0.24\textwidth}
        \includegraphics[width=\linewidth]{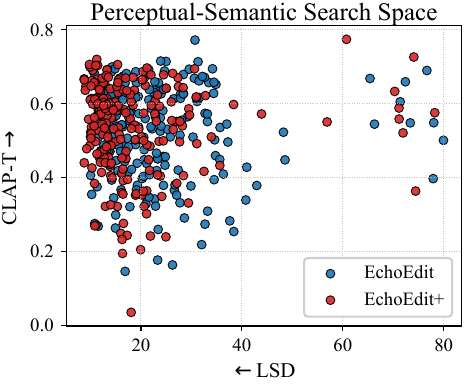}
        \caption{SFX: LSD}
    \end{subfigure}
    \begin{subfigure}[t]{0.24\textwidth}
        \includegraphics[width=\linewidth]{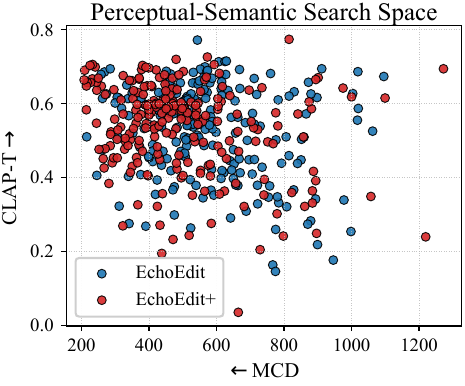}
        \caption{SFX: MCD}
    \end{subfigure}

    \begin{subfigure}[t]{0.24\textwidth}
        \includegraphics[width=\linewidth]{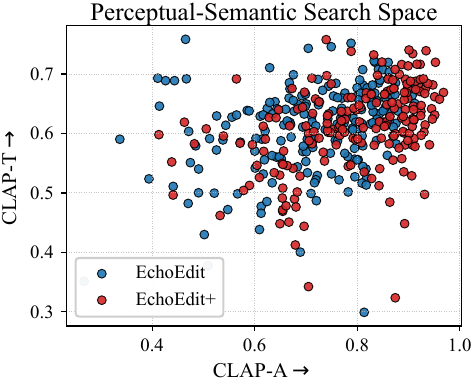}
        \caption{Music: CLAP-A}
    \end{subfigure}
    \begin{subfigure}[t]{0.24\textwidth}
        \includegraphics[width=\linewidth]{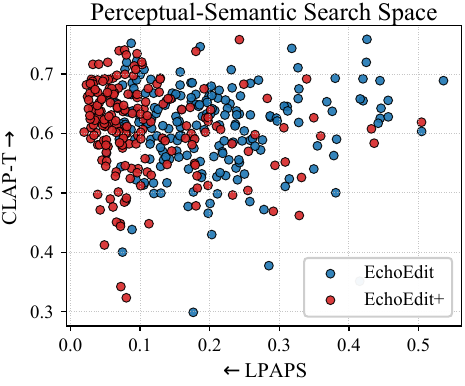}
        \caption{Music: LPAPS}
    \end{subfigure}
    \begin{subfigure}[t]{0.24\textwidth}
        \includegraphics[width=\linewidth]{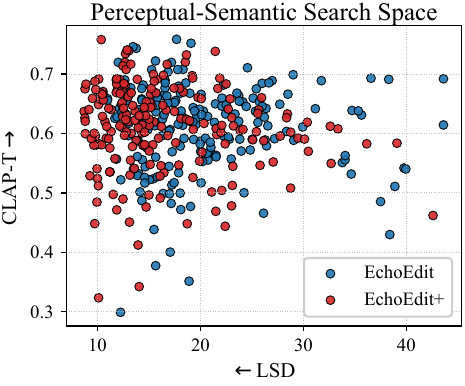}
        \caption{Music: LSD}
    \end{subfigure}
    \begin{subfigure}[t]{0.24\textwidth}
        \includegraphics[width=\linewidth]{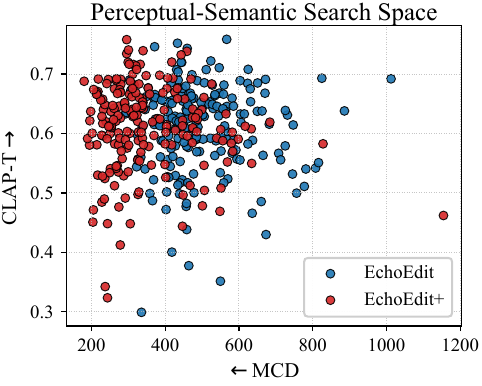}
        \caption{Music: MCD}
    \end{subfigure}
    \caption{\textbf{Raw perceptual--semantic search-space distributions.}
    Each point is an edited example.
    The vertical axis is CLAP-T, and the horizontal axis is a source-preservation or distortion metric.
    \methodplus preserves the target-alignment regime of \method while shifting the population toward stronger preservation and lower acoustic distortion.}
    \label{fig:app_search_space}
\end{figure*}

\begin{figure*}[htpb]
    \centering
    \begin{subfigure}[t]{0.24\textwidth}
        \includegraphics[width=\linewidth]{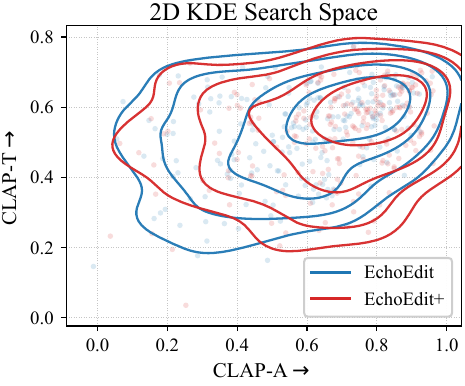}
        \caption{SFX: CLAP-A}
    \end{subfigure}
    \begin{subfigure}[t]{0.24\textwidth}
        \includegraphics[width=\linewidth]{figures/audioedit_comparision/kde_search_space_sfx_lpaps.pdf}
        \caption{SFX: LPAPS}
    \end{subfigure}
    \begin{subfigure}[t]{0.24\textwidth}
        \includegraphics[width=\linewidth]{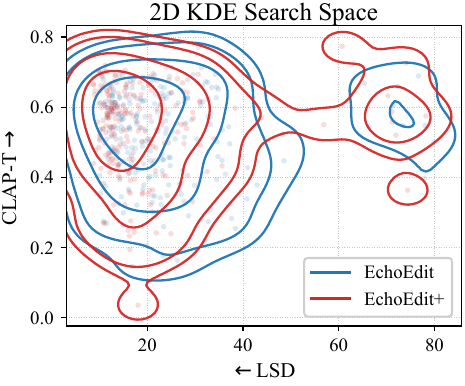}
        \caption{SFX: LSD}
    \end{subfigure}
    \begin{subfigure}[t]{0.24\textwidth}
        \includegraphics[width=\linewidth]{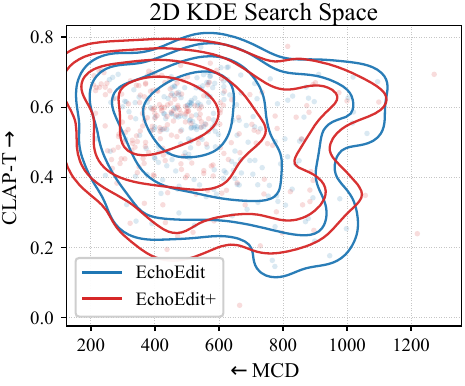}
        \caption{SFX: MCD}
    \end{subfigure}

    \begin{subfigure}[t]{0.24\textwidth}
        \includegraphics[width=\linewidth]{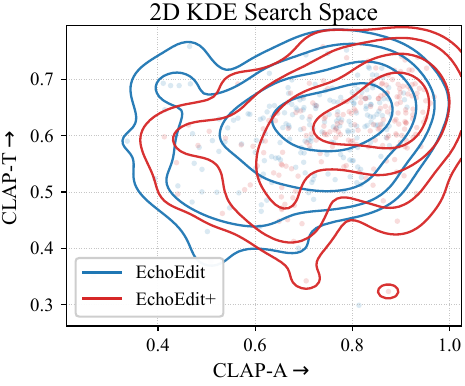}
        \caption{Music: CLAP-A}
    \end{subfigure}
    \begin{subfigure}[t]{0.24\textwidth}
        \includegraphics[width=\linewidth]{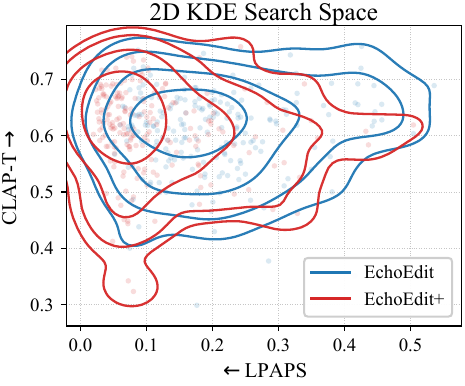}
        \caption{Music: LPAPS}
    \end{subfigure}
    \begin{subfigure}[t]{0.24\textwidth}
        \includegraphics[width=\linewidth]{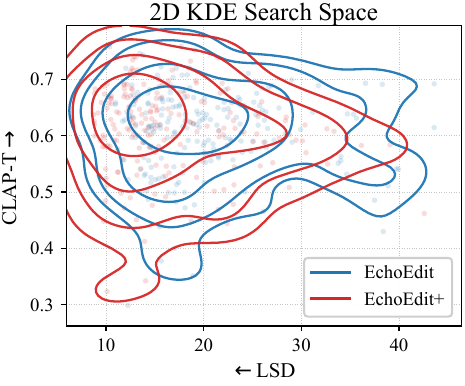}
        \caption{Music: LSD}
    \end{subfigure}
    \begin{subfigure}[t]{0.24\textwidth}
        \includegraphics[width=\linewidth]{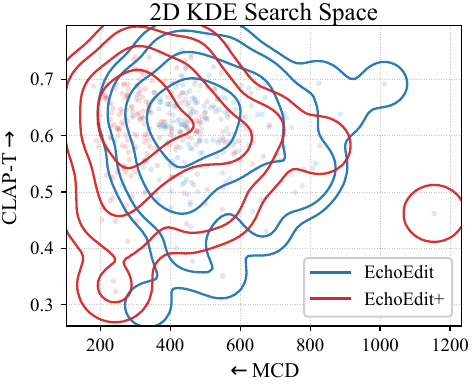}
        \caption{Music: MCD}
    \end{subfigure}
    \caption{\textbf{KDE view of the perceptual--semantic search space.}
    Contours summarize the typical operating region of \method and \methodplus.
    The OT-stabilized editor concentrates probability mass in regions with comparable target alignment and improved preservation, supporting the interpretation that its gain comes from geometric stabilization rather than reduced edit strength.}
    \label{fig:app_kde_space}
\end{figure*}

\begin{figure*}[htpb]
    \centering
    \begin{subfigure}[t]{0.19\textwidth}
        \includegraphics[width=\linewidth]{figures/audioedit_comparision/fsd_sfx_clap_t.pdf}
        \caption{CLAP-T (SFX)}
    \end{subfigure}
    \begin{subfigure}[t]{0.19\textwidth}
        \includegraphics[width=\linewidth]{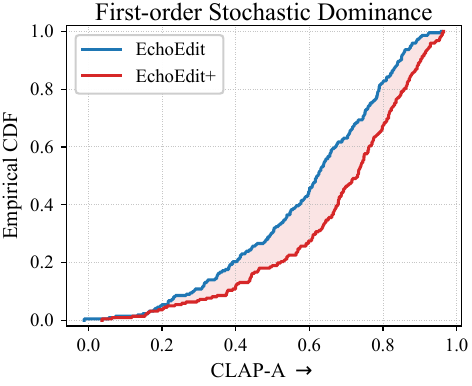}
        \caption{CLAP-A (SFX)}
    \end{subfigure}
    \begin{subfigure}[t]{0.19\textwidth}
        \includegraphics[width=\linewidth]{figures/audioedit_comparision/fsd_sfx_lpaps.pdf}
        \caption{LPAPS (SFX)}
    \end{subfigure}
    \begin{subfigure}[t]{0.19\textwidth}
        \includegraphics[width=\linewidth]{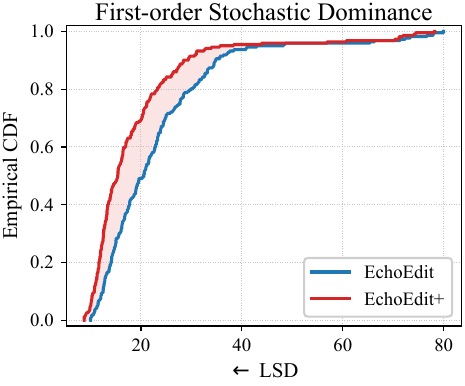}
        \caption{LSD (SFX)}
    \end{subfigure}
    \begin{subfigure}[t]{0.19\textwidth}
        \includegraphics[width=\linewidth]{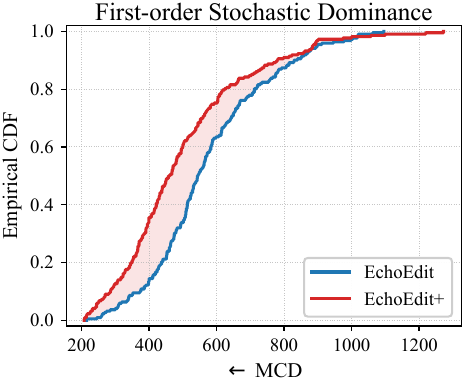}
        \caption{MCD (SFX)}
    \end{subfigure}

    \begin{subfigure}[t]{0.19\textwidth}
        \includegraphics[width=\linewidth]{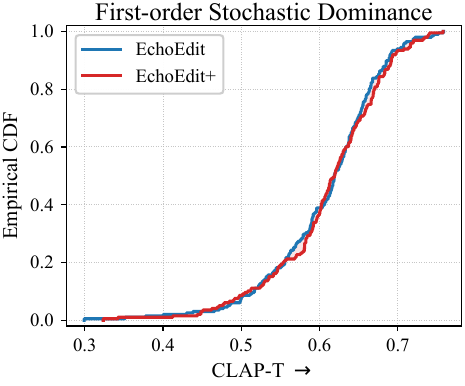}
        \caption{CLAP-T (Music)}
    \end{subfigure}
    \begin{subfigure}[t]{0.19\textwidth}
        \includegraphics[width=\linewidth]{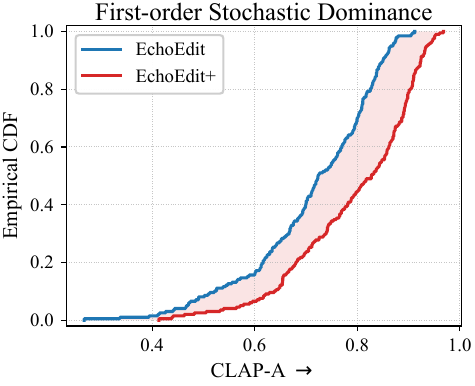}
        \caption{CLAP-A (Music)}
    \end{subfigure}
    \begin{subfigure}[t]{0.19\textwidth}
        \includegraphics[width=\linewidth]{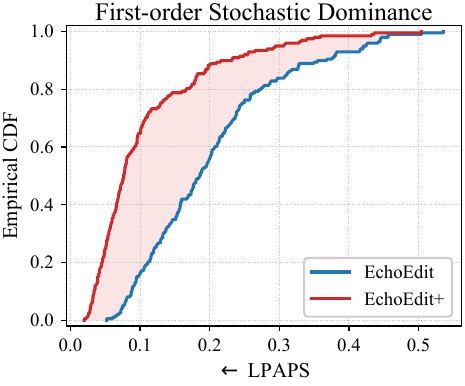}
        \caption{LPAPS (Music)}
    \end{subfigure}
    \begin{subfigure}[t]{0.19\textwidth}
        \includegraphics[width=\linewidth]{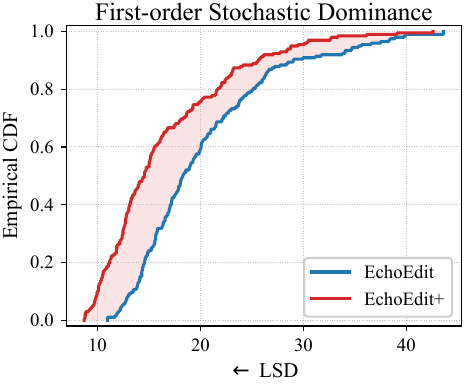}
        \caption{LSD (Music)}
    \end{subfigure}
    \begin{subfigure}[t]{0.19\textwidth}
        \includegraphics[width=\linewidth]{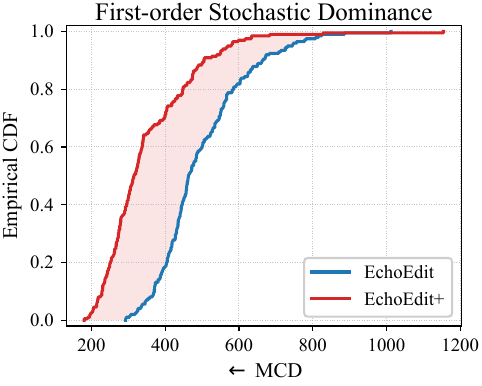}
        \caption{MCD (Music)}
    \end{subfigure}
    \caption{\textbf{Empirical CDFs for sound effects and music.}
    The CDFs compare the full distribution of target alignment and preservation metrics across both domains.
    \methodplus remains close to \method on CLAP-T while improving the distribution of source similarity and distortion-sensitive metrics.
    This supports the claim that OT stabilization preserves long-range audio structure while maintaining effective semantic transport.}
    \label{fig:app_fsd_cdfs}
\end{figure*}

Figure~\ref{fig:app_search_space}--Figure~\ref{fig:app_fsd_cdfs} provide the complete visualization set corresponding to the representative panels in the main paper.
These plots compare \method and \methodplus at the level of distributions rather than only at a single aggregate operating point.
Each search-space panel places target alignment on the vertical axis and a preservation metric on the horizontal axis.
For CLAP-A, larger values indicate stronger source-audio preservation; for LPAPS, LSD, and MCD, smaller values indicate lower perceptual, spectral, or cepstral distortion.
Thus the preferred region is high on the CLAP-T axis and toward the favorable side of the preservation axis.

The raw search-space plots in Figure~\ref{fig:app_search_space} show that OT stabilization does not simply move samples toward weaker edits.
Across both sound effects and music, \methodplus remains in substantially the same CLAP-T regime as \method, while its samples shift toward stronger preservation, especially in LPAPS, LSD, and MCD.
This matches the coupling analysis: replacing the translated stochastic coupling by the OT-admissible coupling should mainly reduce marginal drift in the target-side noisy variable, not erase the conditional velocity residual responsible for semantic change.

Figure~\ref{fig:app_kde_space} smooths the same search-space samples with two-dimensional density contours.
The KDE view is useful because it suppresses isolated outliers and exposes the typical operating region of each method.
For both domains, the high-density region of \methodplus moves toward the favorable preservation side while staying near the high-alignment region.
This density-level behavior is the empirical analogue of the theoretical contraction result: the OT coupling reduces unnecessary noisy-state displacement, so the most probable edits concentrate in a lower-distortion portion of the semantic-preservation space.

The empirical CDFs in Figure~\ref{fig:app_fsd_cdfs} give a complementary one-dimensional view of the same effect.
For target alignment, the CLAP-T curves are close, indicating that \methodplus does not obtain its preservation gains by systematically abandoning the target prompt.
For CLAP-A, the \methodplus distribution shifts toward larger source-audio similarity.
For LPAPS, LSD, and MCD, the \methodplus curves accumulate earlier on the distortion axis, meaning that a larger fraction of examples achieves lower perceptual, spectral, and cepstral change.
The music split exhibits the clearest dominance in the distortion metrics, which is consistent with the main results: musical structure is particularly sensitive to marginal drift, and the OT-admissible coupling provides a strong stabilizing effect.

\subsection{Transport Windows and Style Degrees of Freedom}
\label{app:edit_window_style}

The upper edit-window index determines how much of the direct source-to-target field is traversed before returning to the clean audio endpoint.
In the notation of the derivation, this changes the range of noisy marginals on which the local residual field is queried.
If the OT-admissible coupling has removed the systematic marginal mismatch, increasing this range should expose a controlled frontier: target alignment should strengthen and then saturate, while preservation metrics should change smoothly.
Table~\ref{tab:nmax_app} follows this pattern in both domains.
The sound-effect split shows the expected semantic gain followed by saturation, accompanied by a gradual preservation cost.
The music split exhibits the same semantic trend, while its structural metric remains nearly invariant over the sweep.
Thus $\nmax$ functions as an interpretable transport-length coordinate, not as an uncontrolled perturbation scale.

\begin{table*}[htpb]
    \centering
    \small
    \setlength{\tabcolsep}{4.5pt}
    \begin{tabular}{lccccccc}
        \toprule
        \textbf{Domain} & $\boldsymbol{\nmax}$ & \textbf{CLAP-T} $\uparrow$ & \textbf{CLAP-A} $\uparrow$ & \textbf{LSD} $\downarrow$ & \textbf{MCD} $\downarrow$ & \textbf{LPAPS} $\downarrow$ & \textbf{Structure} $\uparrow$ \\
        \midrule
        \multirow{7}{*}{SFX} & 4  & 0.49 & 0.86 & 11.46 & 222.97 & 0.07 & 0.68 \\
        & 8  & 0.51 & 0.82 & 12.37 & 273.20 & 0.10 & 0.66 \\
        & 12 & 0.52 & 0.75 & 14.72 & 369.19 & 0.13 & 0.62 \\
        & 16 & 0.54 & 0.71 & 17.37 & 452.19 & 0.15 & 0.60 \\
        & 20 & 0.54 & 0.69 & 18.61 & 494.03 & 0.17 & 0.59 \\
        & 24 & 0.53 & 0.68 & 19.01 & 499.67 & 0.17 & 0.59 \\
        & 27 & 0.53 & 0.68 & 19.17 & 505.97 & 0.17 & 0.59 \\
        \midrule
        \multirow{7}{*}{Music} & 4  & 0.58 & 0.84 & 10.41 & 161.80 & 0.04 & 0.99 \\
        & 8  & 0.59 & 0.84 & 10.72 & 176.77 & 0.05 & 0.99 \\
        & 12 & 0.60 & 0.82 & 12.04 & 230.92 & 0.07 & 0.99 \\
        & 16 & 0.61 & 0.81 & 14.84 & 309.47 & 0.09 & 0.98 \\
        & 20 & 0.61 & 0.80 & 15.88 & 336.60 & 0.10 & 0.98 \\
        & 24 & 0.61 & 0.79 & 16.38 & 350.08 & 0.11 & 0.98 \\
        & 27 & 0.61 & 0.79 & 16.57 & 353.76 & 0.11 & 0.98 \\
        \bottomrule
    \end{tabular}
    \caption{Effect of the edit-window upper index $\nmax$. Larger windows trace a controlled semantic--preservation frontier rather than an unstable perturbation regime.}
    \label{tab:nmax_app}
\end{table*}

The lower edit index governs a complementary degree of freedom.
It determines how much of the late, low-noise part of the trajectory is assigned to direct OT-regularized editing rather than target-conditioned completion.
These low-noise states carry detailed phase, timbre, event placement, and musical contour; consequently, excessive target-only freedom should weaken the source-conditioned manifold constraint.
Table~\ref{tab:nmin_app} verifies this qualitative behavior.
A small positive lower index can be beneficial for style-oriented prompts, where the target asks for broad timbral or production changes.
Beyond that regime, preservation degrades coherently across acoustic and structural metrics.
This is the empirical counterpart of the theoretical trade-off: style freedom can be useful, but only while the edited path remains close to the local source-conditioned manifold.

\begin{table*}[htpb]
    \centering
    \small
    \setlength{\tabcolsep}{4.5pt}
    \begin{tabular}{lccccccc}
        \toprule
        \textbf{Domain} & $\boldsymbol{\nmin}$ & \textbf{CLAP-T} $\uparrow$ & \textbf{CLAP-A} $\uparrow$ & \textbf{LSD} $\downarrow$ & \textbf{MCD} $\downarrow$ & \textbf{LPAPS} $\downarrow$ & \textbf{Structure} $\uparrow$ \\
        \midrule
        \multirow{5}{*}{SFX} & 0  & 0.53 & 0.68 & 19.00 & 499.62 & 0.17 & 0.59 \\
        & 4  & 0.54 & 0.68 & 17.95 & 486.85 & 0.16 & 0.59 \\
        & 8  & 0.50 & 0.59 & 19.12 & 543.61 & 0.21 & 0.52 \\
        & 12 & 0.46 & 0.49 & 23.41 & 621.77 & 0.26 & 0.49 \\
        & 16 & 0.43 & 0.43 & 29.11 & 684.59 & 0.29 & 0.47 \\
        \midrule
        \multirow{5}{*}{Music} & 0  & 0.61 & 0.80 & 16.38 & 350.12 & 0.11 & 0.98 \\
        & 4  & 0.62 & 0.81 & 15.15 & 328.09 & 0.10 & 0.98 \\
        & 8  & 0.60 & 0.77 & 15.32 & 365.07 & 0.14 & 0.98 \\
        & 12 & 0.57 & 0.70 & 18.46 & 505.76 & 0.22 & 0.94 \\
        & 16 & 0.55 & 0.63 & 22.71 & 606.82 & 0.26 & 0.86 \\
        \bottomrule
    \end{tabular}
    \caption{Effect of the style-control lower index $\nmin$. Moderate low-noise freedom can strengthen style transfer, whereas excessive freedom weakens source-conditioned consistency.}
    \label{tab:nmin_app}
\end{table*}

The dedicated style-edit setting in Table~\ref{tab:style_app} isolates the same phenomenon.
\methodplus does not attempt to maximize semantic displacement at all costs.
Instead, it selects a lower-cost admissible displacement that remains compatible with the target condition.
This yields a consistent preservation advantage over the original stochastic \method, especially in metrics sensitive to acoustic identity and musical structure.
The sound-effect case also illustrates an important nuance: when the style prompt demands a broad timbral shift, OT regularization may trade a small amount of target-score aggressiveness for a more coherent source-preserving edit.

\begin{table*}[htpb]
    \centering
    \small
    \setlength{\tabcolsep}{4.5pt}
    \begin{tabular}{llcccccc}
        \toprule
        \textbf{Domain} & \textbf{Method} & \textbf{CLAP-T} $\uparrow$ & \textbf{CLAP-A} $\uparrow$ & \textbf{LSD} $\downarrow$ & \textbf{MCD} $\downarrow$ & \textbf{LPAPS} $\downarrow$ & \textbf{Structure} $\uparrow$ \\
        \midrule
        \multirow{2}{*}{SFX} & \method & 0.51 & 0.34 & 22.98 & 649.63 & 0.36 & 0.46 \\
        & \methodplus & 0.46 & 0.45 & 17.11 & 498.13 & 0.26 & 0.54 \\
        \midrule
        \multirow{2}{*}{Music} & \method & 0.58 & 0.66 & 21.69 & 561.32 & 0.24 & 0.94 \\
        & \methodplus & 0.58 & 0.77 & 16.05 & 377.23 & 0.12 & 0.97 \\
        \bottomrule
    \end{tabular}
    \caption{Dedicated style-edit setting. \methodplus emphasizes lower-cost source-preserving style transport while retaining competitive target alignment.}
    \label{tab:style_app}
\end{table*}

\subsection{Stochastic Stability and Solver Calibration}
\label{app:solver_calibration}

The residual analysis separates two error sources: stochastic measurement noise and coupling-induced marginal drift.
This separation yields a direct empirical test.
If the OT correction removes the systematic drift term, then increasing the number of stochastic velocity samples should provide incremental refinement rather than a qualitatively different editing regime.
Table~\ref{tab:navg_app} is consistent with this prediction.
With the higher-resolution solver, a single velocity sample already lies on the same semantic--preservation frontier as larger averages.
Averaging becomes more useful when the solver is coarser, where it compensates for a rougher local field estimate.
Thus Monte Carlo averaging improves numerical smoothness, but the dominant stabilization mechanism is the admissible coupling itself.

\begin{table*}[htpb]
    \centering
    \small
    \setlength{\tabcolsep}{4.5pt}
    \begin{tabular}{lcccccccc}
        \toprule
        \textbf{Domain} & $\boldsymbol{T}$ & $\boldsymbol{\navg}$ & \textbf{CLAP-T} $\uparrow$ & \textbf{CLAP-A} $\uparrow$ & \textbf{LSD} $\downarrow$ & \textbf{MCD} $\downarrow$ & \textbf{LPAPS} $\downarrow$ & \textbf{Structure} $\uparrow$ \\
        \midrule
        \multirow{8}{*}{SFX} & \multirow{4}{*}{10} & 1  & 0.51 & 0.64 & 19.55 & 537.50 & 0.18 & 0.57 \\
        & & 3  & 0.52 & 0.68 & 19.09 & 509.24 & 0.17 & 0.59 \\
        & & 5  & 0.52 & 0.68 & 19.03 & 502.81 & 0.16 & 0.59 \\
        & & 10 & 0.51 & 0.69 & 18.72 & 496.07 & 0.15 & 0.59 \\
        \cmidrule(l){2-9}
        & \multirow{4}{*}{28} & 1  & 0.53 & 0.68 & 19.01 & 499.68 & 0.17 & 0.59 \\
        & & 3  & 0.53 & 0.69 & 18.66 & 494.13 & 0.15 & 0.59 \\
        & & 5  & 0.52 & 0.69 & 18.55 & 493.26 & 0.15 & 0.60 \\
        & & 10 & 0.52 & 0.69 & 18.54 & 493.12 & 0.15 & 0.60 \\
        \midrule
        \multirow{8}{*}{Music} & \multirow{4}{*}{10} & 1  & 0.60 & 0.76 & 17.85 & 399.90 & 0.14 & 0.98 \\
        & & 3  & 0.61 & 0.80 & 16.54 & 355.71 & 0.10 & 0.98 \\
        & & 5  & 0.61 & 0.80 & 16.47 & 348.33 & 0.10 & 0.98 \\
        & & 10 & 0.61 & 0.80 & 16.23 & 344.34 & 0.10 & 0.98 \\
        \cmidrule(l){2-9}
        & \multirow{4}{*}{28} & 1  & 0.61 & 0.80 & 16.38 & 350.10 & 0.11 & 0.98 \\
        & & 3  & 0.61 & 0.80 & 15.89 & 336.00 & 0.09 & 0.98 \\
        & & 5  & 0.61 & 0.80 & 15.65 & 327.98 & 0.09 & 0.98 \\
        & & 10 & 0.61 & 0.80 & 15.56 & 325.12 & 0.09 & 0.98 \\
        \bottomrule
    \end{tabular}
    \caption{Velocity averaging under 10-step and 28-step solvers. Once the OT coupling is used, averaging mainly provides secondary variance reduction.}
    \label{tab:navg_app}
\end{table*}

The step coefficient $\eta$ probes whether the estimated residual field has a stable numerical scale.
If the field were dominated by random directionality, changing $\eta$ would produce irregular behavior.
Table~\ref{tab:eta_app} instead shows an ordered response.
Small coefficients under-apply the target edit while preserving the source; overly large coefficients overshoot the low-cost path and increase distortion.
The default region lies near the knee of this frontier, where target alignment has largely saturated but preservation has not yet deteriorated sharply.
This behavior supports the view that \methodplus exposes a calibrated transport direction rather than an arbitrary noise-amplification knob.

\begin{table*}[htpb]
    \centering
    \small
    \setlength{\tabcolsep}{4.5pt}
    \begin{tabular}{lccccccc}
        \toprule
        \textbf{Domain} & $\boldsymbol{\eta}$ & \textbf{CLAP-T} $\uparrow$ & \textbf{CLAP-A} $\uparrow$ & \textbf{LSD} $\downarrow$ & \textbf{MCD} $\downarrow$ & \textbf{LPAPS} $\downarrow$ & \textbf{Structure} $\uparrow$ \\
        \midrule
        \multirow{7}{*}{SFX} & 0.4 & 0.50 & 0.83 & 13.03 & 288.85 & 0.08 & 0.66 \\
        & 0.6 & 0.52 & 0.79 & 14.86 & 361.28 & 0.11 & 0.64 \\
        & 0.8 & 0.53 & 0.73 & 16.93 & 434.01 & 0.14 & 0.61 \\
        & 1.0 & 0.53 & 0.68 & 19.00 & 499.64 & 0.17 & 0.59 \\
        & 1.2 & 0.53 & 0.63 & 21.12 & 558.13 & 0.19 & 0.57 \\
        & 1.4 & 0.52 & 0.59 & 23.16 & 609.23 & 0.21 & 0.55 \\
        & 1.6 & 0.51 & 0.56 & 25.16 & 651.27 & 0.23 & 0.53 \\
        \midrule
        \multirow{7}{*}{Music} & 0.4 & 0.59 & 0.84 & 11.88 & 198.37 & 0.05 & 0.99 \\
        & 0.6 & 0.60 & 0.83 & 13.13 & 239.10 & 0.06 & 0.99 \\
        & 0.8 & 0.61 & 0.82 & 14.65 & 290.25 & 0.08 & 0.99 \\
        & 1.0 & 0.61 & 0.80 & 16.38 & 350.09 & 0.11 & 0.98 \\
        & 1.2 & 0.61 & 0.77 & 18.31 & 415.76 & 0.13 & 0.98 \\
        & 1.4 & 0.60 & 0.74 & 20.35 & 480.95 & 0.17 & 0.97 \\
        & 1.6 & 0.58 & 0.69 & 22.36 & 542.06 & 0.20 & 0.97 \\
        \bottomrule
    \end{tabular}
    \caption{Step coefficient $\eta$. The sweep reveals a calibrated transport scale with a clear semantic--preservation knee.}
    \label{tab:eta_app}
\end{table*}

\subsection{Operation-Type and Prompt Robustness}
\label{app:operation_robustness}

A source-preserving editor must remain stable under qualitatively different edit operations.
Replacement changes the identity of an existing event, addition introduces a new attribute, and deletion suppresses source content while preserving the surrounding scene or musical context.
These operations stress different regions of the source-conditioned target law.
Table~\ref{tab:operation_app} shows that the OT-stabilized editor remains consistently strong across this breakdown.
\methodplus is best or tied-best on target alignment and dominates the preservation-sensitive columns in both domains.
Deletion is especially informative because it can easily degenerate into unrestricted regeneration; the improved preservation of \methodplus indicates that removing a target attribute does not require abandoning the source trajectory.
The operation-wise evidence therefore supports a broad stabilization claim rather than a category-specific improvement.

\begin{table*}[htpb]
    \centering
    \small
    \setlength{\tabcolsep}{4.5pt}
    \begin{tabular}{lllcccccc}
        \toprule
        \textbf{Domain} & \textbf{Operation} & \textbf{Method} & \textbf{CLAP-T} $\uparrow$ & \textbf{CLAP-A} $\uparrow$ & \textbf{LSD} $\downarrow$ & \textbf{MCD} $\downarrow$ & \textbf{LPAPS} $\downarrow$ & \textbf{Structure} $\uparrow$ \\
        \midrule
        \multirow{15}{*}{SFX} & \multirow{5}{*}{replacement} & FireFlow & 0.38 & 0.35 & 30.12 & 686.54 & 0.33 & 0.45 \\
        & & ODE Inv. & 0.42 & 0.37 & 28.42 & 676.01 & 0.29 & 0.48 \\
        & & SDEdit & 0.42 & 0.38 & 26.90 & 667.53 & 0.31 & 0.46 \\
        & & \method & \textbf{0.53} & 0.52 & 23.70 & 591.37 & 0.25 & 0.54 \\
        & & \textbf{\methodplus} & \textbf{0.53} & \textbf{0.60} & \textbf{19.94} & \textbf{536.89} & \textbf{0.20} & \textbf{0.57} \\
        \cmidrule(l){2-9}
        & \multirow{5}{*}{addition} & FireFlow & 0.42 & 0.44 & 27.44 & 667.87 & 0.29 & 0.48 \\
        & & ODE Inv. & 0.46 & 0.46 & 28.00 & 689.25 & 0.27 & 0.49 \\
        & & SDEdit & 0.44 & 0.47 & 26.97 & 657.69 & 0.28 & 0.48 \\
        & & \method & 0.53 & 0.62 & 24.07 & 567.84 & 0.21 & 0.55 \\
        & & \textbf{\methodplus} & \textbf{0.54} & \textbf{0.70} & \textbf{19.96} & \textbf{492.68} & \textbf{0.17} & \textbf{0.58} \\
        \cmidrule(l){2-9}
        & \multirow{5}{*}{deletion} & FireFlow & 0.28 & 0.29 & 32.80 & 703.35 & 0.33 & 0.43 \\
        & & ODE Inv. & 0.34 & 0.36 & 30.38 & 700.31 & 0.31 & 0.47 \\
        & & SDEdit & 0.46 & 0.52 & 24.68 & 627.30 & 0.24 & 0.51 \\
        & & \method & 0.53 & 0.64 & 22.19 & 567.86 & 0.19 & 0.58 \\
        & & \textbf{\methodplus} & \textbf{0.54} & \textbf{0.74} & \textbf{17.04} & \textbf{468.38} & \textbf{0.13} & \textbf{0.62} \\
        \midrule
        \multirow{15}{*}{Music} & \multirow{5}{*}{replacement} & FireFlow & 0.56 & 0.62 & 24.51 & 631.84 & 0.30 & 0.87 \\
        & & ODE Inv. & 0.58 & 0.64 & 23.29 & 631.29 & 0.28 & 0.89 \\
        & & SDEdit & 0.54 & 0.64 & 20.49 & 555.38 & 0.28 & 0.90 \\
        & & \method & \textbf{0.61} & 0.67 & 20.95 & 514.06 & 0.25 & 0.91 \\
        & & \textbf{\methodplus} & \textbf{0.61} & \textbf{0.76} & \textbf{17.54} & \textbf{387.62} & \textbf{0.14} & \textbf{0.98} \\
        \cmidrule(l){2-9}
        & \multirow{5}{*}{addition} & FireFlow & 0.57 & 0.69 & 23.89 & 592.89 & 0.23 & 0.89 \\
        & & ODE Inv. & 0.59 & 0.71 & 23.54 & 590.29 & 0.22 & 0.91 \\
        & & SDEdit & 0.57 & 0.68 & 20.46 & 543.18 & 0.22 & 0.89 \\
        & & \method & 0.61 & 0.74 & 21.09 & 507.92 & 0.19 & 0.91 \\
        & & \textbf{\methodplus} & \textbf{0.62} & \textbf{0.80} & \textbf{17.26} & \textbf{366.47} & \textbf{0.10} & \textbf{0.98} \\
        \cmidrule(l){2-9}
        & \multirow{5}{*}{deletion} & FireFlow & 0.57 & 0.66 & 23.67 & 606.45 & 0.23 & 0.90 \\
        & & ODE Inv. & 0.59 & 0.67 & 22.63 & 599.68 & 0.21 & 0.91 \\
        & & SDEdit & 0.57 & 0.69 & 19.15 & 530.35 & 0.20 & 0.93 \\
        & & \method & 0.60 & 0.73 & 18.74 & 470.73 & 0.16 & 0.93 \\
        & & \textbf{\methodplus} & \textbf{0.61} & \textbf{0.82} & \textbf{14.33} & \textbf{296.22} & \textbf{0.08} & \textbf{0.98} \\
        \bottomrule
    \end{tabular}
    \caption{Operation-type breakdown. The OT-stabilized editor preserves its advantage across replacement, addition, and deletion edits.}
    \label{tab:operation_app}
\end{table*}

Because the method compares source- and target-conditioned velocities, one might expect it to depend strongly on the precision of the source text.
The sensitivity study in Table~\ref{tab:robust_app} argues against this failure mode.
Empty, metadata-derived, and paraphrased source prompts lead to nearly identical trade-offs, and the seed variation remains small.
This is consistent with the coupling analysis.
The source audio itself supplies the local geometric anchor, while the OT regularizer keeps the edited noisy marginal aligned with that anchor even when the linguistic source description is incomplete.

\begin{table*}[htpb]
    \centering
    \small
    \setlength{\tabcolsep}{4.5pt}
    \begin{tabular}{llcccccc}
        \toprule
        \textbf{Domain} & \textbf{Setting} & \textbf{CLAP-T} $\uparrow$ & \textbf{CLAP-A} $\uparrow$ & \textbf{LSD} $\downarrow$ & \textbf{MCD} $\downarrow$ & \textbf{LPAPS} $\downarrow$ & \textbf{Structure} $\uparrow$ \\
        \midrule
        \multirow{3}{*}{SFX} & empty source prompt & 0.53 & 0.66 & 18.28 & 566.01 & 0.17 & 0.60 \\
        & metadata prompt & 0.52 & 0.67 & 16.46 & 506.93 & 0.18 & 0.61 \\
        & paraphrase prompt & 0.51 & 0.65 & 16.98 & 509.88 & 0.17 & 0.61 \\
        \cmidrule(lr){1-8}
        \multirow{3}{*}{Music} & empty source prompt & 0.58 & 0.76 & 15.84 & 391.14 & 0.11 & 0.96 \\
        & metadata prompt & 0.58 & 0.78 & 14.82 & 353.33 & 0.10 & 0.97 \\
        & paraphrase prompt & 0.58 & 0.78 & 15.15 & 370.49 & 0.11 & 0.97 \\
        \midrule
        SFX & seed std. & 0.00 & 0.01 & 0.06 & 4.26 & 0.00 & 0.00 \\
        Music & seed std. & 0.00 & 0.00 & 0.27 & 3.11 & 0.00 & 0.00 \\
        \bottomrule
    \end{tabular}
    \caption{Source-prompt and seed robustness. Source-prompt rows are means over the sensitivity subset; seed rows report standard deviation over five seeds.}
    \label{tab:robust_app}
\end{table*}

\subsection{Sequential Editing and Accumulated Drift}
\label{app:chained_editing}

Sequential editing distinguishes local consistency from global semantic accumulation.
After each edit, the source-conditioned target law for the next stage is different; therefore, no local OT argument can guarantee indefinite closeness to the original recording.
Table~\ref{tab:chained_app} makes this distinction explicit by reporting preservation against both the initial source and the immediately preceding stage.
The source-relative metrics naturally degrade as semantically different edits accumulate.
In contrast, the previous-stage metrics remain substantially more stable, especially for music.
This is precisely the behavior expected from a local transport regularizer: \methodplus stabilizes each step of the chain, while the chain as a whole still reflects the semantic history of the requested edits.

\begin{table*}[htpb]
    \centering
    \small
    \setlength{\tabcolsep}{4.5pt}
    \begin{tabular}{lccccccc}
        \toprule
        \textbf{Domain} & \textbf{Stage} & \textbf{CLAP-T} $\uparrow$ & \textbf{CLAP-A Src} $\uparrow$ & \textbf{LPAPS Src} $\downarrow$ & \textbf{LPAPS Prev} $\downarrow$ & \textbf{Structure Src} $\uparrow$ & \textbf{Structure Prev} $\uparrow$ \\
        \midrule
        \multirow{5}{*}{SFX} & 0 & 0.53 & 0.72 & 0.11 & 0.11 & 0.65 & 0.65 \\
        & 1 & 0.47 & 0.59 & 0.19 & 0.17 & 0.60 & 0.64 \\
        & 3 & 0.47 & 0.38 & 0.34 & 0.22 & 0.45 & 0.60 \\
        & 5 & 0.39 & 0.41 & 0.27 & 0.12 & 0.47 & 0.64 \\
        & 7 & 0.35 & 0.33 & 0.27 & 0.10 & 0.47 & 0.66 \\
        \midrule
        \multirow{5}{*}{Music} & 0 & 0.67 & 0.89 & 0.08 & 0.08 & 0.99 & 0.99 \\
        & 1 & 0.63 & 0.84 & 0.14 & 0.07 & 0.97 & 0.99 \\
        & 3 & 0.59 & 0.72 & 0.25 & 0.11 & 0.98 & 0.99 \\
        & 5 & 0.52 & 0.62 & 0.31 & 0.08 & 0.95 & 1.00 \\
        & 7 & 0.40 & 0.58 & 0.35 & 0.05 & 0.92 & 0.99 \\
        \bottomrule
    \end{tabular}
    \caption{Sequential editing. ``Src'' compares each stage to the original source; ``Prev'' compares it to the immediately preceding stage.}
    \label{tab:chained_app}
\end{table*}

\subsection{Spectral Structure of the Direct Editing Field}
\label{app:velocity_spectra}

\begin{figure*}[htpb]
    \centering
    \begin{subfigure}[t]{0.47\textwidth}
        \centering
        \includegraphics[width=\linewidth]{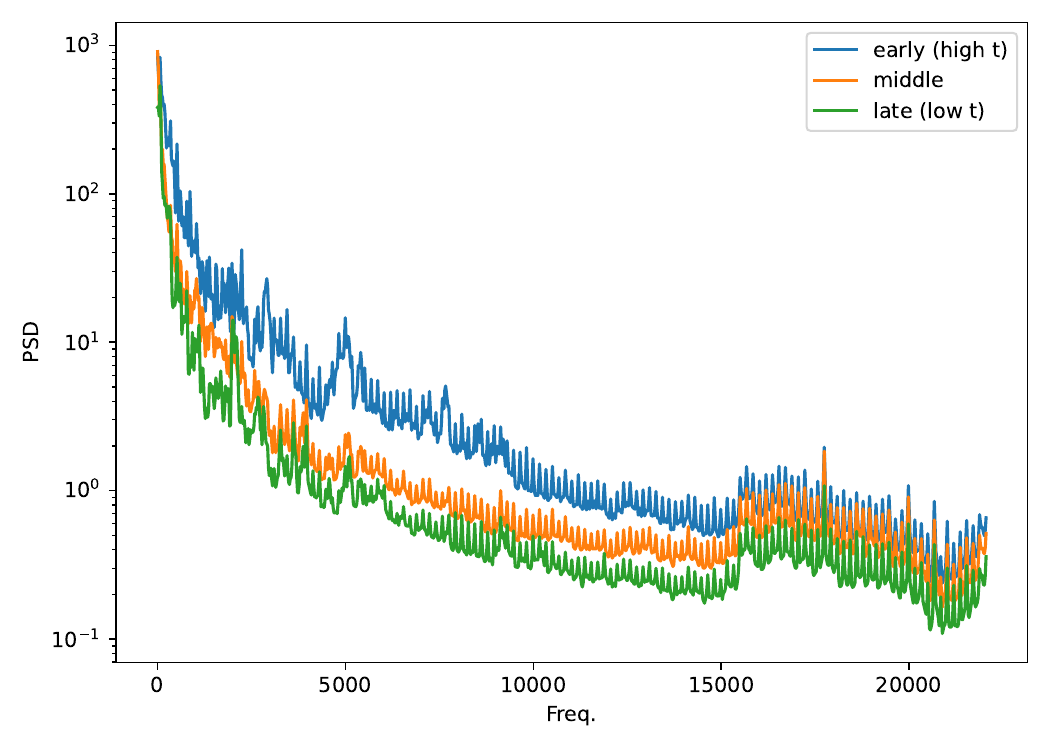}
        \caption{SFX: Velocity-difference power spectra.}
        \label{fig:sfx_lines}
    \end{subfigure}
    \hfill
    \begin{subfigure}[t]{0.47\textwidth}
        \centering
        \includegraphics[width=\linewidth]{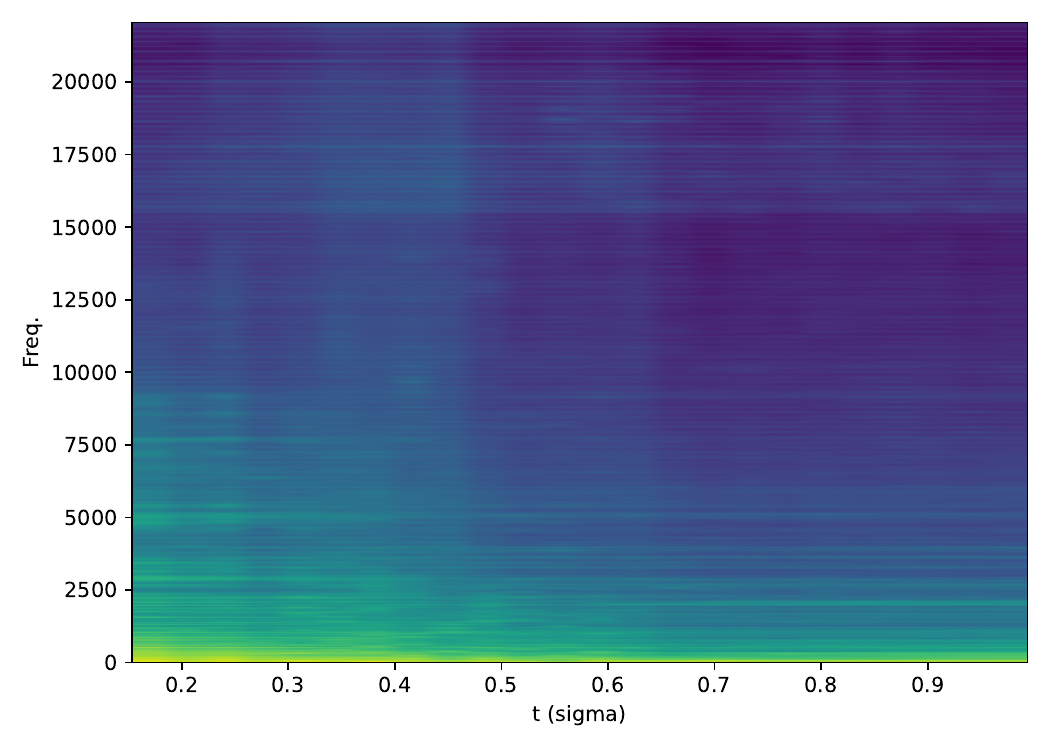}
        \caption{SFX: Power-spectral heatmap over flow noise levels.}
        \label{fig:sfx_heatmap}
    \end{subfigure}

    \begin{subfigure}[t]{0.47\textwidth}
        \centering
        \includegraphics[width=\linewidth]{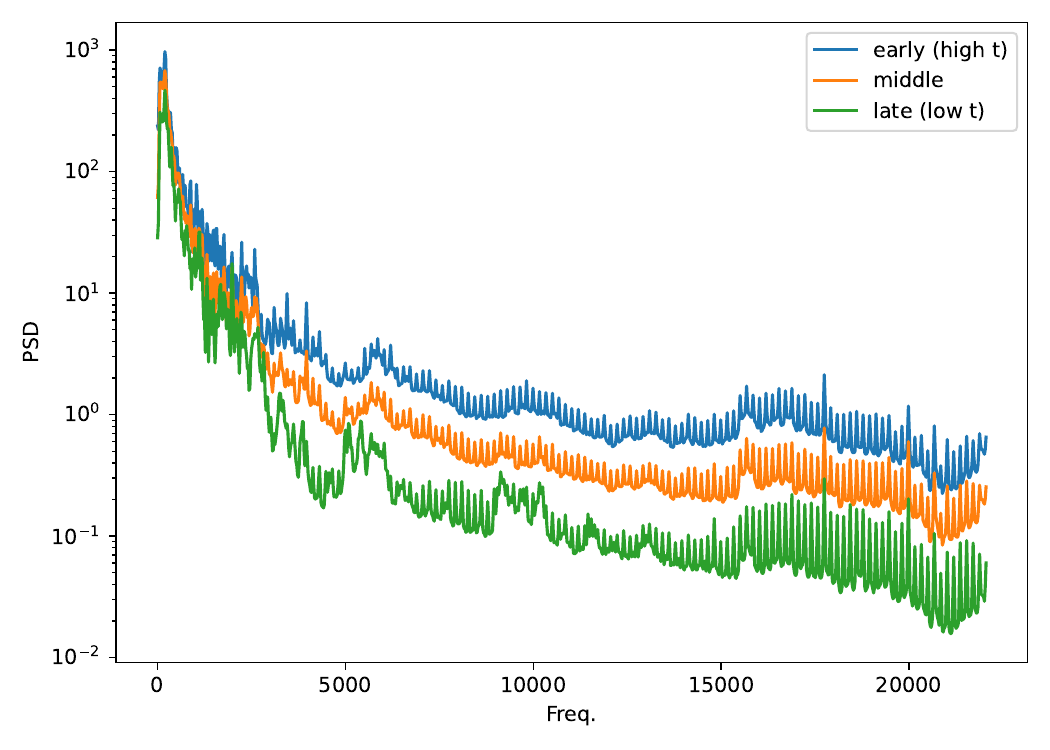}
        \caption{Music: Velocity-difference power spectra.}
        \label{fig:music_lines}
    \end{subfigure}
    \hfill
    \begin{subfigure}[t]{0.47\textwidth}
        \centering
        \includegraphics[width=\linewidth]{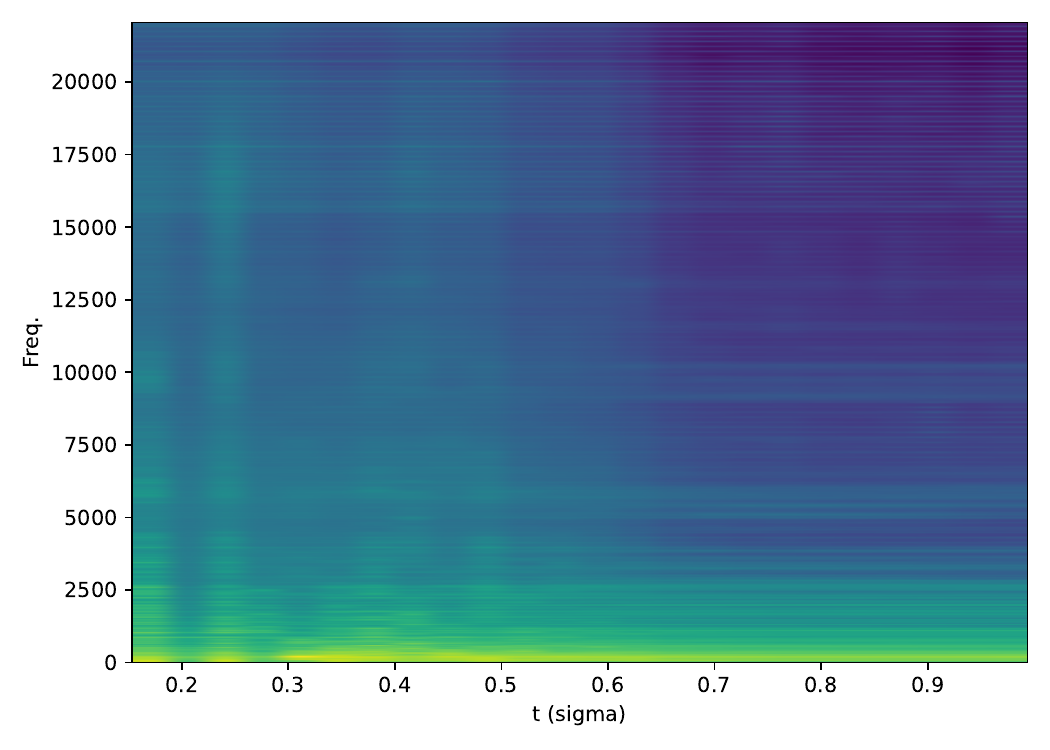}
        \caption{Music: Power-spectral heatmap over flow noise levels.}
        \label{fig:music_heatmap}
    \end{subfigure}
    \caption{\textbf{Decoded velocity-difference spectral analysis.} The residual field is structured across flow noise levels instead of behaving as broadband stochastic corruption.}
    \label{fig:velocity_spectra_combined}
\end{figure*}

The direct residual field should have acoustic structure if it is induced by the learned audio-flow geometry.
If it were merely an unconstrained stochastic perturbation, decoded residual energy would be broadband and inconsistent across flow noise levels.
Figure~\ref{fig:velocity_spectra_combined} shows the opposite behavior.
Across sound effects and music, the residual power concentrates in low- and mid-frequency regions, with substantially less high-frequency energy.
This supports the manifold interpretation: the field modifies global acoustic identity, rhythm, and timbral envelope while avoiding the broadband corruption expected from unregularized noise injection.

Table~\ref{tab:psd_band_app} provides a compact band-averaged summary of the same trend.
The low- and mid-frequency bands dominate across representative noise levels, whereas the highest band remains comparatively small.
The result complements the residual-bias analysis.
OT controls the noisy marginal at which the residual is evaluated, and the measured residual retains coherent acoustic structure rather than collapsing into high-frequency artifacts.

\begin{table*}[htpb]
    \centering
    \small
    \setlength{\tabcolsep}{4.5pt}
    \begin{tabular}{lcccccc}
        \toprule
        \textbf{Domain} & \textbf{Noise level} & \textbf{0--250 Hz} & \textbf{250--1k Hz} & \textbf{1--4k Hz} & \textbf{4--8k Hz} & \textbf{8--16k Hz} \\
        \midrule
        \multirow{3}{*}{SFX} & 0.99 & 218.26 & 49.70 & 16.84 & 4.88 & 1.35 \\
        & 0.82 & 331.90 & 33.37 & 6.93 & 2.14 & 0.74 \\
        & 0.15 & 312.88 & 31.68 & 6.14 & 1.76 & 0.57 \\
        \midrule
        \multirow{3}{*}{Music} & 0.99 & 412.80 & 63.94 & 11.87 & 2.44 & 0.85 \\
        & 0.82 & 439.15 & 55.69 & 6.23 & 0.70 & 0.26 \\
        & 0.15 & 292.91 & 39.80 & 4.04 & 0.31 & 0.11 \\
        \bottomrule
    \end{tabular}
    \caption{Band-averaged decoded velocity-difference power at representative noise levels. Energy concentration in low- and mid-frequency bands indicates a structured audio residual.}
    \label{tab:psd_band_app}
\end{table*}

\subsection{Duration Scaling and Boundary Regimes}
\label{app:duration_limitations}

Long-form music editing tests whether the local transport field remains usable as the latent sequence becomes substantially longer.
Table~\ref{tab:duration_app} shows no evidence of metric collapse across the evaluated duration range.
The preservation-sensitive metrics remain stable while target alignment stays within the same operating regime as shorter clips.
This does not imply that every long edit is perceptually flawless; long clips can still reveal localized vocal instability or incomplete arrangement changes.
It does indicate, however, that the OT-regularized direct field scales to the practical duration range of the backbone without the numerical fragility commonly associated with accumulated inversion error.

\begin{table}[t]
    \centering
    \small
    \setlength{\tabcolsep}{2.1pt}
    \begin{tabular}{@{}cccccc@{}}
        \toprule
        \textbf{Seconds} & \textbf{CLAP-T} $\uparrow$ & \textbf{CLAP-A} $\uparrow$ & \textbf{LSD} $\downarrow$ & \textbf{LPAPS} $\downarrow$ & \textbf{Structure} $\uparrow$ \\
        \midrule
        15  & 0.53 & 0.74 & 16.64 & 0.13 & 1.00 \\
        30  & 0.57 & 0.80 & 14.19 & 0.11 & 0.99 \\
        45  & 0.59 & 0.83 & 12.45 & 0.09 & 0.99 \\
        60  & 0.60 & 0.83 & 11.84 & 0.09 & 0.99 \\
        75  & 0.60 & 0.87 & 10.95 & 0.07 & 0.99 \\
        90  & 0.60 & 0.87 & 11.24 & 0.07 & 0.99 \\
        105 & 0.59 & 0.87 & 11.10 & 0.07 & 0.99 \\
        120 & 0.60 & 0.86 & 11.11 & 0.08 & 0.99 \\
        \bottomrule
    \end{tabular}
    \caption{Music duration scaling. The OT-regularized direct field remains stable across the evaluated long-form range.}
    \label{tab:duration_app}
\end{table}

\paragraph{Boundary cases.}
\label{app:limitation_subset}
The difficult cases involve acoustically incompatible replacements, structurally complex mixtures, vocal-to-instrumental transformations, solo-to-ensemble changes, and extreme genre shifts.
Such prompts weaken the assumption that the edit can be represented as a local source-conditioned transport.
The resulting failures are therefore theoretically meaningful rather than accidental.
\methodplus can regularize a trajectory toward a source-conditioned audio manifold, but it cannot make that manifold well defined when the prompt effectively asks for a new scene, a new arrangement, or a different musical object.
In these regimes, aggregate metrics may still show partial preservation, while perceptual inspection reveals source leakage, incomplete semantic replacement, or localized instability.


\section{User Study}
\label{app:user_study}

\begin{figure*}[htpb]
    \centering
    \includegraphics[width=0.74\textwidth,height=0.78\textheight,keepaspectratio]{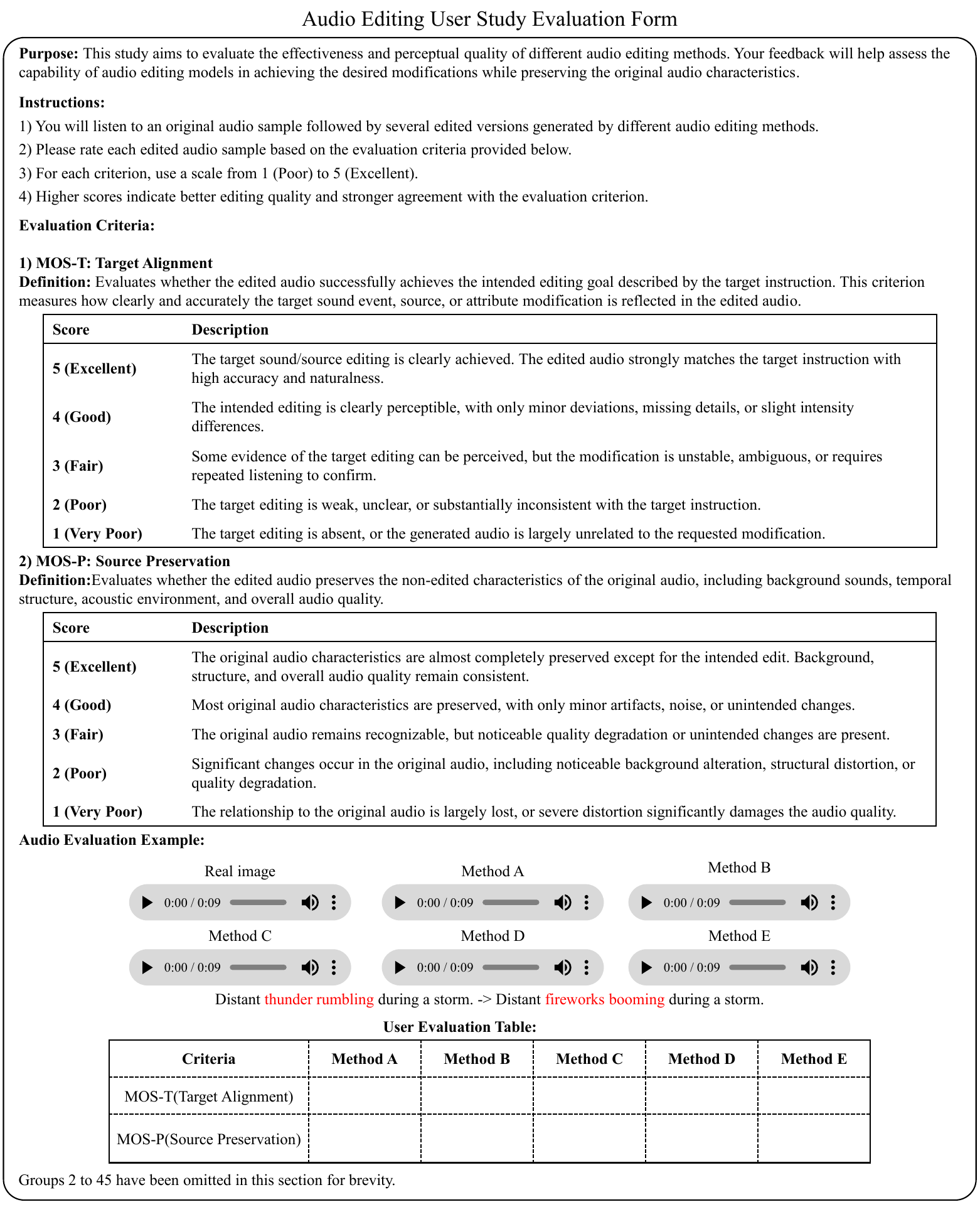}
    \caption{\textbf{MOS evaluation form for audio editing.}
    Each item presents the source audio, the source-to-target prompt pair, and anonymized edited outputs from the evaluated methods.
    Listeners assign independent 1--5 scores for target alignment and source preservation, with higher scores indicating stronger perceptual agreement with the criterion.}
    \label{fig:app_mos_form}
\end{figure*}

To complement the objective metrics, we conduct a mean-opinion-score study that directly evaluates whether listeners perceive an edit as prompt-faithful and source-preserving.
The study follows the same motivation as the subjective evaluation used in prior editing work: automatic metrics summarize particular acoustic or embedding distances, but they do not fully capture naturalness, transient stability, musical continuity, or the practical usefulness of the edited result.
An example of the evaluation form is shown in Figure~\ref{fig:app_mos_form}.

\paragraph{Protocol.}
Each evaluation item contains the original audio, a source-to-target prompt pair, and five edited outputs produced by FireFlow, ODE inversion, SDEdit, \method, and \methodplus.
The method identities are hidden from listeners.
For each edited output, listeners provide two scores on a 1--5 Likert scale.
\textbf{MOS-T} measures whether the requested target event, instrument, style, or acoustic attribute is clearly realized.
\textbf{MOS-P} measures whether the non-edited content of the source remains perceptually intact, including timing, background texture, acoustic environment, rhythm, melody, and overall audio quality.
We report \textbf{Overall} as the arithmetic mean of MOS-T and MOS-P, so that the summary score reflects the central semantic-preservation trade-off rather than an additional independent judgment.
The final aggregation contains 2,250 scored method instances from ten completed annotation sheets, with 450 ratings per method and 750 ratings per edit operation.


\begin{table}[t]
    \centering
    \small
    \setlength{\tabcolsep}{4.5pt}
    \begin{tabular}{lccc}
        \toprule
        \textbf{Method} & \textbf{MOS-T} $\uparrow$ & \textbf{MOS-P} $\uparrow$ & \textbf{Overall} $\uparrow$ \\
        \midrule
        FireFlow & 2.822 & 2.676 & 2.749 \\
        ODE Inv. & 2.958 & 2.836 & 2.897 \\
        SDEdit & 3.276 & 3.109 & 3.192 \\
        \method & 4.358 & 4.467 & 4.412 \\
        \textbf{\methodplus} & \textbf{4.393} & \textbf{4.580} & \textbf{4.487} \\
        \bottomrule
    \end{tabular}
    \caption{Aggregate MOS results. Each method receives 450 ratings. \methodplus achieves the best target alignment, source preservation, and overall perceptual quality.}
    \label{tab:app_mos_method}
\end{table}

Table~\ref{tab:app_mos_method} shows a clear perceptual separation between direct editing and inversion/noising baselines.
SDEdit is the strongest baseline among the three inversion-style methods, but it remains substantially below \method and \methodplus on both perceptual axes.
This gap is consistent with the qualitative evidence: a noisy reconstruction bridge can make the target concept audible, but it often damages event placement, background texture, or musical continuity.
The comparison between \method and \methodplus is more diagnostic.
Both methods are perceived as strongly target-aligned, but \methodplus obtains the larger gain on MOS-P, improving source preservation by 0.113 while also improving overall MOS by 0.075.
This supports the interpretation that the OT-admissible coupling does not merely intensify the semantic edit; it regularizes the trajectory so that listeners perceive fewer unintended changes.

\begin{table}[t]
    \centering
    \small
    \setlength{\tabcolsep}{4.5pt}
    \begin{tabular}{llccc}
        \toprule
        \textbf{Operation} & \textbf{Method} & \textbf{MOS-T} $\uparrow$ & \textbf{MOS-P} $\uparrow$ & \textbf{Overall} $\uparrow$ \\
        \midrule
        \multirow{5}{*}{Addition} & FireFlow & 2.847 & 2.660 & 2.753 \\
        & ODE Inv. & 2.913 & 2.720 & 2.817 \\
        & SDEdit & 3.253 & 3.087 & 3.170 \\
        & \method & \textbf{4.353} & 4.407 & 4.380 \\
        & \methodplus & 4.300 & \textbf{4.533} & \textbf{4.417} \\
        \midrule
        \multirow{5}{*}{Deletion} & FireFlow & 2.880 & 2.747 & 2.813 \\
        & ODE Inv. & 3.033 & 2.893 & 2.963 \\
        & SDEdit & 3.493 & 3.267 & 3.380 \\
        & \method & 4.460 & 4.553 & 4.507 \\
        & \methodplus & \textbf{4.607} & \textbf{4.713} & \textbf{4.660} \\
        \midrule
        \multirow{5}{*}{Replacement} & FireFlow & 2.740 & 2.620 & 2.680 \\
        & ODE Inv. & 2.927 & 2.893 & 2.910 \\
        & SDEdit & 3.080 & 2.973 & 3.027 \\
        & \method & 4.260 & 4.440 & 4.350 \\
        & \methodplus & \textbf{4.273} & \textbf{4.493} & \textbf{4.383} \\
        \bottomrule
    \end{tabular}
    \caption{MOS breakdown by edit operation. Each method-operation cell contains 150 ratings. Best values within each operation are bold.}
    \label{tab:app_mos_operation}
\end{table}

The operation-level results in Table~\ref{tab:app_mos_operation} further clarify the perceptual role of OT stabilization.
Averaged over all methods, deletion receives the highest overall MOS (3.665), followed by addition (3.507) and replacement (3.470), reflecting that replacing an existing acoustic identity is often the most ambiguous source-conditioned transport problem.
Within each operation, however, the relative pattern remains stable: \methodplus obtains the best overall MOS for addition, deletion, and replacement.
The only exception is MOS-T for addition, where \method is slightly higher than \methodplus.
This exception is informative rather than contradictory: direct stochastic transport can sometimes sound marginally more aggressive when introducing new content, while the OT correction favors edits that listeners judge as better preserved and more coherent overall.
Thus, the subjective study reinforces the central claim of the paper: \methodplus improves the perceptual quality of inversion-free audio editing primarily by suppressing uncertainty drift, not by weakening or replacing the semantic editing field.

\end{document}